\documentclass[aps,prx,twocolumn,superscriptaddress,nobibnotes,a4paper,super]{revtex4-2}
\usepackage{algorithm}
\usepackage{algpseudocode}
\usepackage[utf8]{inputenc}
\usepackage{mathptmx} 
\usepackage{amsmath}
\usepackage{graphicx,bm,hyperref,siunitx,cleveref,url}

\hypersetup{
	colorlinks   = true, 
	urlcolor     = blue, 
	linkcolor    = blue, 
	citecolor   = red 
}
\usepackage{siunitx} 
\usepackage{soul} 
\begin{document}
\setcitestyle{super} 

\title{Centimeter-scale nanomechanical resonators with low dissipation}

\author{Andrea Cupertino}\thanks{These authors contributed equally: Andrea Cupertino, Dongil Shin.}
\affiliation{Department of Precision and Microsystems Engineering, Delft University of Technology, Mekelweg 2, 2628 CD, Delft, The Netherlands}

\author{Dongil Shin}\thanks{These authors contributed equally: Andrea Cupertino, Dongil Shin.}
\affiliation{Department of Precision and Microsystems Engineering, Delft University of Technology, Mekelweg 2, 2628 CD, Delft, The Netherlands}
\affiliation{Department of Materials Science and Engineering, Delft University of Technology, Mekelweg 2, 2628 CD, Delft, The Netherlands}

\author{Leo Guo}
\affiliation{Department of Microelectronics, Delft University of Technology, Mekelweg 2, 2628 CD, Delft, The Netherlands}

\author{Peter G. Steeneken}
\affiliation{Department of Precision and Microsystems Engineering, Delft University of Technology, Mekelweg 2, 2628 CD, Delft, The Netherlands}
\affiliation{Kavli Institute of Nanoscience, Department of Quantum Nanoscience, Delft University of Technology, Lorentzweg 1, 2628 CJ, Delft, The Netherlands}

\author{Miguel A. Bessa}
\email[email:]{miguel\_bessa@brown.edu}
\affiliation{School of Engineering, Brown University, 184 Hope St., Providence, RI, 02912, USA}

\author{Richard A. Norte}
\email[email:]{r.a.norte@tudelft.nl}
\affiliation{Department of Precision and Microsystems Engineering, Delft University of Technology, Mekelweg 2, 2628 CD, Delft, The Netherlands}
\affiliation{Kavli Institute of Nanoscience, Department of Quantum Nanoscience, Delft University of Technology, Lorentzweg 1, 2628 CJ, Delft, The Netherlands}

\begin{abstract}
High-aspect-ratio mechanical resonators are pivotal in precision sensing, from macroscopic gravitational wave detectors to nanoscale acoustics. However, fabrication challenges and high computational costs have limited the length-to-thickness ratio of these devices, leaving a largely unexplored regime in nano-engineering. We present nanomechanical resonators that extend centimeters in length yet retain nanometer thickness. We explore this expanded design space using an optimization approach which judiciously employs fast millimeter-scale simulations to steer the more computationally intensive centimeter-scale design optimization. By employing delicate nanofabrication techniques, our approach ensures high-yield realization, experimentally confirming room-temperature quality factors close to theoretical predictions. The synergy between nanofabrication, design optimization guided by machine learning, and precision engineering opens a solid-state path to room-temperature quality factors approaching 10 billion at kilohertz mechanical frequencies -- comparable to the performance of leading cryogenic resonators and levitated nanospheres, even under significantly less stringent temperature and vacuum conditions.
\end{abstract}

\maketitle

\section*{INTRODUCTION}

Mechanical resonators are crucial in precision sensing, enabling gravitational-wave observations at the macroscale\cite{abramoviciLIGOLaserInterferometer1992,gonzalezBrownianMotionMass1994}, probing weak forces in atomic force microscopy at the nanoscale \cite{sidlesMagneticResonanceForce1995, binnigAtomicForceMicroscope1986, rossiVectorialScanningForce2017, nicholNanomechanicalDetectionNuclear2012, harrisQuantumMechanicalSystems2015} or playing a central role in recent quantum technologies\cite{guoFeedbackCoolingRoom2019, masonContinuousForceDisplacement2019, saarinenLaserCoolingMembraneinthemiddle2023}. Their performance largely hinges on having low mechanical dissipation, quantified by the mechanical quality factor (Q). The Q factor measures both radiated acoustic energy and infiltrating thermomechanical noise, with a high Q pivotal in preserving resonator coherence. This is crucial, particularly at room temperature, for observing quantum phenomena\cite{seisGroundStateCooling2022a,midoloNanooptoelectromechanicalSystems2018}, advancing quantum technology\cite{magriniRealtimeOptimalQuantum2021}, and maximizing sensitivity for detecting changes in mass\cite{manzanequeMethodDetermineClosedLoop2020, chenDiamagneticallyLevitatingResonant2021, hanayInertialImagingNanomechanical2015}, force\cite{reinhardt2016ultralow, eichlerUltrahighQNanomechanicalResonators2022}, and displacement\cite{abramoviciLIGOLaserInterferometer1992}. High Q is often achieved through ``dissipation dilution", a phenomenon originating from the synergistic effects of large tensile stress and high-aspect-ratio, observed both in resonators with macroscopic lengths on the order of centimeters and above\cite{gonzalezBrownianMotionMass1994, cagnoliDampingDilutionFactor2000} and resonators with nanometers thicknesses\cite{unterreithmeierDampingNanomechanicalResonators2010, schmidDampingMechanismsHighQ2011, yuControlMaterialDamping2012}. The reader is referred to Cagnoli et al.\cite{cagnoliDampingDilutionFactor2000} for a discussion about the use of `dissipation dilution' terminology for pendula and prestressed nanomechanical resonators.

\begin{figure*} 
   	\centering
	\includegraphics[width = 1\textwidth]{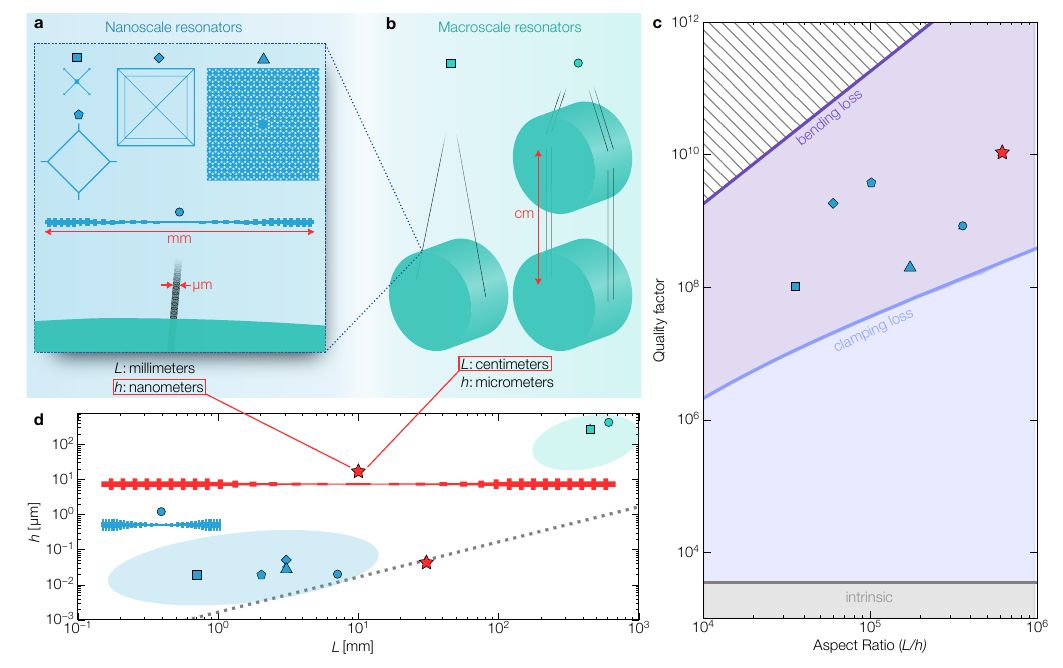} 
	\caption{\textbf{High-aspect-ratio mechanical resonators}. \textbf{a}, State-of-the-art Si\textsubscript{3}N\textsubscript{4} nanomechanical resonators with nanometers thickness and length below 1 centimeter. On the left side: trampoline resonator \cite{norteMechanicalResonatorsQuantum2016} and perimeter resonator \cite{bereyhiPerimeterModesNanomechanical2022}. At the center: spiderweb resonator \cite{shinSpiderwebNanomechanicalResonators2022}. On the right side: soft-clamped membrane \cite{tsaturyanUltracoherentNanomechanicalResonators2017}. On the bottom: tapered string with phononic crystal \cite{ghadimiElasticStrainEngineering2018} and a zoom-in of the LIGO suspension in \textbf{b}, scaled to match the dimension of all the nanomechanical resonators. \textbf{b}, Macromechanical resonators with micrometers thickness and extending into tens of centimeters. From left to right: LIGO suspension \cite{dawidInvestigationViolinMode1997}, advanced LIGO suspension \cite{cummingDesignDevelopmentAdvanced2012}. \textbf{c}, Predicted quality factor as a function of the mechanical resonators aspect-ratio for a Si\textsubscript{3}N\textsubscript{4} string. The blue points indicate the measured quality factor for the devices in \textbf{a}. The red stars show the predicted quality factor of the design demonstrated here. \textbf{d}, Thickness versus length of state-of-the-art mechanical resonators in \textbf{a} and \textbf{b}. The red star shows the design demonstrated here with its aspect-ratio indicated by the dotted black line.}
	\label{fig:1_Intro}
\end{figure*}

At the macroscale, a notable example is the pendulum formed by the mirror-suspension pair in gravitation wave detectors \cite{cummingDesignDevelopmentAdvanced2012, dawidInvestigationViolinMode1997}, where kilogram mirrors are suspended by wires tens of centimeters long and a thickness on the order of micrometers (Fig. \ref{fig:1_Intro}b). The high tensile stress is created by the mirror masses, which induce stress in the wires due to gravity. The resulting high quality factor of the pendulum modes of the order of $10^8$ allows isolating the detector from thermomechanical noise, enabling it to reach its enhanced displacement sensitivity. The exceptional acoustic isolation of these resonators has enabled landmark demonstrations of quantum effects at an unprecedented kilogram scale \cite{whittleApproachingMotionalGround2021a} and contributed to the first observations of gravitational waves \cite{abbott2016observation}. Similar configurations at smaller scales are employed on table-top experiments to investigate the limits of quantum mechanics \cite{corbittOpticalDilutionFeedback2007} and its interplay with gravity \cite{matsumotoDemonstrationDisplacementSensing2019}. To date, resonators in this category, which span lengths of several centimeters, have been limited to minimum thicknesses on the order of micrometers. These types of mechanical resonators have been particularly successful in demonstrating low dissipation in room-temperature environments.

At the nanoscale, nanomechanical resonators possess significantly reduced thickness, in the range of sub-micrometers, with a length limited to a sub-millimeter length by convention. They are generally fabricated on top of a supporting substrate by standard silicon technology and integrated onto a single chip (Fig. \ref{fig:1_Intro}a). High tensile stress, generated through a thermal mismatch between the resonator and the substrate, offers both lower mechanical dissipation and structural stability for long-distance suspended nanostructures. This necessitates careful material selection. Silicon nitride (Si\textsubscript{3}N\textsubscript{4}) is emerging as one of the most common and easily manufacturable materials. It has favorable optical and mechanical properties at room temperature and specific advantages gained through strained configurations on silicon, emphasizing the role of strain engineering for dissipation characteristics. Leveraging standard silicon technology compatibility, these nanomechanical resonators provide scalability\cite{westerveld2021sensitive}, and integration\cite{guoFeedbackCoolingRoom2019} not found in their macroscale counterparts. Techniques like mode-shape engineering\cite{norteMechanicalResonatorsQuantum2016, bereyhiHierarchicalTensileStructures2022, prattNanoscaleTorsionalDissipation2023,bereyhiPerimeterModesNanomechanical2022, shinSpiderwebNanomechanicalResonators2022} and phononic crystal (PnC) engineering\cite{tsaturyanUltracoherentNanomechanicalResonators2017,ghadimiElasticStrainEngineering2018} have further enhanced dissipation dilution, pushing quality factors above $10^9$. This improved coherence of motion has propelled advances in quantum phenomena exploration at cryogenic temperatures \cite{rossiMeasurementbasedQuantumControl2018, purdyObservationRadiationPressure2013} and fostered an advanced sensor class development\cite{halgMembraneBasedScanningForce2021, chasteNanomechanicalMassSensor2012}. Generally, the quality factor in this class of resonators (Fig. \ref{fig:1_Intro}d) is proportional to the length, and the resonance frequency is inversely proportional to the length \cite{schmidDampingMechanismsHighQ2011,shinSpiderwebNanomechanicalResonators2022}.
As nanoscale resonators become longer, their manufacturing yield drops dramatically due to compounding factors including limits in devices per chip, the alignment of nanoscale features over centimeters, and as resonators become longer they become increasingly delicate suspended structures. These combined decrease device yield and experimental match with expected Qs as resonators become more susceptible to small forces during nanofabrication. These overlapping fabrication challenges have made reliably increasing the length of nanoscale resonators beyond millimeter scales prohibitively difficult in practice.

Advancing these resonators to multi-centimeter lengths while maintaining their nanoscale thickness would uniquely combine the benefits of macroscale and nanoscale mechanical resonators and open an undiscovered regime for acoustic technologies. These suspended structures will be characterized by their ultrahigh room-temperature Q factors, their ability to firmly integrate with microchip architectures, and their relatively large masses and surface areas. Such centimeter-scale surface areas and masses which are mechanically well-isolated are well suited for high precision measurements of acceleration\cite{krause2012high}, pressure\cite{naserbakht2019squeeze}, and vacuum\cite{scherschligt2018quantum}. These attributes also make them promising for the observation of mesoscopic quantum behavior in ambient temperatures which are largely limited by room-temperature thermomechanical noise\cite{huang2023room}. Reaching a quantum-limited motion regime at room temperature will extend the application range of quantum sensors\cite{magriniRealtimeOptimalQuantum2021}, quantum calibrated sensing \cite{purdyQuantumCorrelationsRoomtemperature2017} and quantum memories \cite{kristensenLonglivedEfficientOptomechanical2023}.The drive towards multi-centimeter nanostructures also holds the potential for applications in ultrasensitive nanomechanical detection\cite{zhangZeptonewtonForceSensing2022},
including searches for dark-matter\cite{manleySearchingVectorDark2021}, Casimir forces\cite{pate2020casimir, fong2019phonon}, and studies of entropy and time\cite{pearson2021measuring}. Yet, pursuing such high-aspect-ratios at the envisioned centimeter scale faces considerable computational and fabrication barriers, including the need for a high fabrication yield to offset the limited number of devices per chip and cost implications. This challenge of cost-effective high-yield fabrication for centimeter-scale nanostructures introduces a specific dynamic between economics and design distinct from traditional high-volume nanotechnology.

We propose an approach that merges delicate fabrication techniques and multi-fidelity Bayesian optimization, enabling the creation of mechanical resonators with centimeter lengths and high-aspect-ratios exceeding $4.3 \times 10^5$; equivalent to reliably producing ceramic structures with a thickness of 1 millimeter, suspended over nearly half a kilometer. Our design strategy curtails optimization time while successfully circumventing common fabrication issues such as stiction, collapse and fracture. Applying our strategy to a \SI{3}{cm} long silicon nitride string, we achieve a quality factor exceeding $6.5 \times 10^9$ at room temperature -- the highest ever recorded for a mechanically clamped resonator. This solid-state platform performs on par with counterparts such as levitated nanospheres which require significantly more stringent vacuum conditions as low as $10^{-11}$ mbar\cite{daniaUltrahighQualityFactor2023} to reduce the collision rate between the particle and background gas molecules, typically the dominant source of dissipation. In contrast, our clamped centimeter-scale resonators are limited by intrinsic losses \cite{unterreithmeierDampingNanomechanicalResonators2010} in view of their higher vibrational frequency, and can approach comparable quality factors at pressures nearly two orders of magnitude higher. Notably, our room-temperature resonators can also work at quality factors only observed in cryogenic counterparts \cite{beccariStrainedCrystallineNanomechanical2022, maccabe2020nano}. This enhanced capability to operate at higher temperatures and pressures unveils the potentials in centimeter-scale nanotechnology, expanding the boundaries of what is achievable with on-chip, room-temperature resonators.

\section*{RESULTS}

\subsection*{High-aspect-ratio advantage and multi-fidelity design}
\label{sec:2_Qequation}

\begin{figure*} 
    \centering
	\includegraphics[width = 1.0\textwidth]{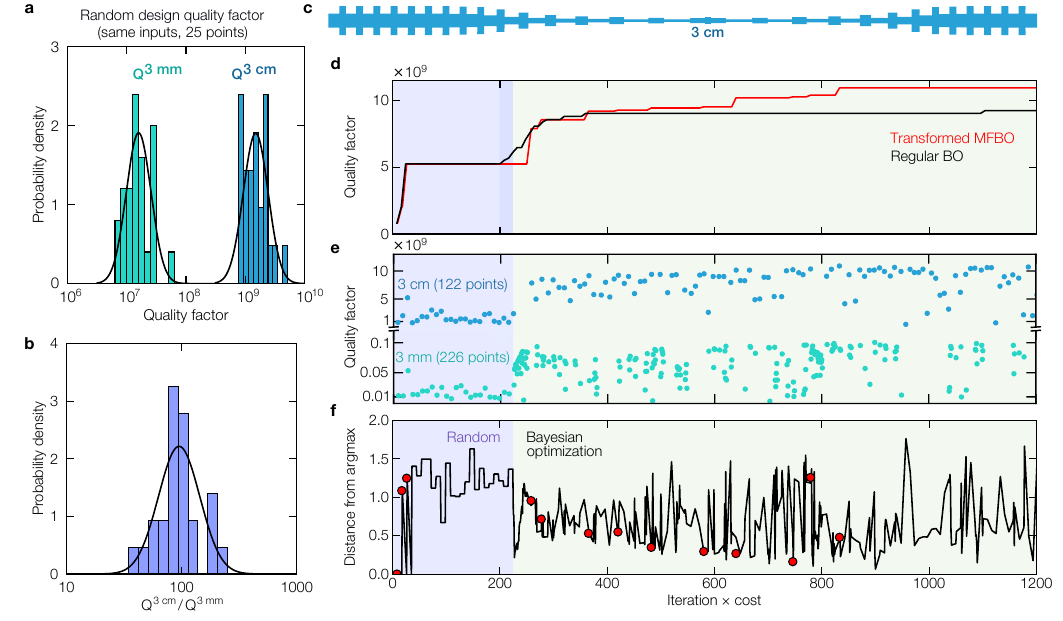} 
	\caption{
 \textbf{Multi-fidelity Bayesian optimization}
 \textbf{a}, Quality factor distribution for the two different lengths of PnC resonators (Blue: \SI{3}{cm}, Green: \SI{3}{mm}). 25 sets of design parameters were randomly selected, and both results followed a log-normal distribution.
 \textbf{b}, Probability distribution of the \SI{3}{cm} and \SI{3}{mm} resonator's quality factor ratio. The quality factor ratio follows a log-normal distribution.
 \textbf{c}, Shape of the optimized \SI{3}{cm} resonator \textbf{d}, Evolution of the optimized quality factor with two formulations. MFBO maximizing the log of the quality factor (Transformed MFBO) and single-fidelity Bayesian optimization maximizing the quality factor directly (Regular BO). Transformed MFBO outperformed regular BO.
 \textbf{e}, Iteration history of the transformed MFBO where the shaded blue part represents the first 25 randomly selected simulations for both high and low-fidelity models, and where the green part represents the remaining design iterations where the MFBO method searches the best design and controls what fidelity it wants to evaluate. The abscissa of the plot has units of iteration cost, and its limit was set to 1200. At the end of the optimization process, 122 high-fidelity simulations (each with a relative cost of 8) and 226 low-fidelity simulations (relative cost of 1) were evaluated. All the designs considered during the optimization can be found in the Supplementary Video.
 \textbf{f}, The distance from a previous optimized point to the point considered in that iteration. The red markers indicate when the algorithm has found a higher quality factor until that iteration.
 \textbf{d} to \textbf{f} shares the same x-axis corresponding to the design iteration times the cost of each iteration as the optimization process evolves.}
	\label{fig:2_Design}
\end{figure*}

The quality factor of string resonators with a constant cross-section is given by \cite{schmidDampingMechanismsHighQ2011}

\begin{equation}
	\label{eq:Qtot}
	Q = Q_{\rm int} \Big[n^2\pi^2\frac{E}{12\sigma}\Big(\frac{h}{L}\Big)^2 + 2\sqrt{\frac{E}{12\sigma}}\Big(\frac{h}{L}\Big)\Big]^{-1}
\end{equation}
where $Q_{\rm int}$ is the intrinsic quality factor (surface loss) that varies linearly with the resonator’s thickness, $n$ is the mode order, $E$ is the Young’s modulus, $\sigma$ is the initial stress, $L$ and $h$ are the resonator's length and thickness in the direction of motion, respectively.
Note that this equation assumes perfect clamping to the substrate without considering the mechanical coupling, potentially affecting the Q factor at a smaller mode order~\cite{de2022mechanical}.
Since $Q_{\rm int} \propto h$, increasing the length can be a more effective strategy (than reducing thickness) to increase quality factors. For given aspect ratio \textit{h}/\textit{L}, the first and second term in the denominator of \autoref{eq:Qtot} originates from bending loss and clamping loss, respectively, for which we use the definitions from Schmid \textit{et al.}~\cite{schmidDampingMechanismsHighQ2011}.
The clamping loss originates from the sharp curvature at the clamps where the resonator is anchored to the supporting substrate, and the bending loss accounts for the curvature along the remainder of the resonator, both depending on the aspect-ratio (Fig. \ref{fig:1_Intro}c).
The clamping region is defined as the part of the resonator at a distance of less than $L_{\rm c}=\sqrt{E/6\sigma} h$ from the clamping points where the mode shape deviates strongly from a simple sine function~\cite{schmidDampingMechanismsHighQ2011}. The curvature is defined as the reciprocal of the radius of curvature for the resonator's bending deformation.

In the high-aspect-ratio limit, the contribution of clamping loss is orders of magnitude larger, dominating the loss contribution. To this end, techniques known as soft clamping have been employed to reduce the sharp curvature at this $L_{\rm c}$ region \cite{ghadimiElasticStrainEngineering2018} and improve the Q factor so that the Q factor scales quadratically with the aspect-ratio of the resonator. Those include phononic crystal-based string and membrane resonators, which employ a higher-order eigenmode confined in a central defect by the acoustic bandgap\cite{tsaturyanUltracoherentNanomechanicalResonators2017,ghadimiElasticStrainEngineering2018}, spiderweb and perimeters resonators, which exploit low-order eigenmodes\cite{shinSpiderwebNanomechanicalResonators2022, bereyhiPerimeterModesNanomechanical2022}, or other methods considering the fundamental mode\cite{norteMechanicalResonatorsQuantum2016,bereyhiHierarchicalTensileStructures2022} (Fig. \ref{fig:1_Intro}a). It is important to notice that \autoref{eq:Qtot} is a straight beam formulation that does not directly apply to phononic crystal strings \cite{fedorov2019generalized} (due to the geometry difference) but can give insight into general trends with $h/L$. A high aspect-ratio is beneficial even if the Q factor is limited by the curvature at the clamping points (\autoref{eq:Qtot}). Hence, state-of-the-art high Q factor nanoresonators have advanced towards devices with increasing aspect-ratios, pushing the total length from micrometers \cite{thompson2008strong, norteMechanicalResonatorsQuantum2016} to millimeters \cite{shinSpiderwebNanomechanicalResonators2022, tsaturyanUltracoherentNanomechanicalResonators2017, ghadimiElasticStrainEngineering2018, bereyhiPerimeterModesNanomechanical2022} (Fig. \ref{fig:1_Intro}c) alongside a thickness reduction below \SI{50}{nm}.

Centimeter-long resonators maintaining a thickness within tens of nanometers can significantly raise the achievable quality factor.
However, the resulting higher aspect-ratios demand advanced simulation-based design strategies due to the elevated computational cost of accurately capturing their behavior through direct simulation models. Numerical analyses require increased degrees of freedom (DOFs) to describe models with extreme aspect ratios. For instance, conducting finite element analysis on a centimeter-long string with phononic crystals requires roughly ten times more DOFs than the same geometry with a one order lower aspect-ratio. This poses a challenge for centimeter-scale designs, as the heightened computational cost sets a practical boundary, particularly given the limited availability of high-fidelity data.
One optimization process took about 16 CPU-days for 150 iterations, which created a bottleneck in the design process. 
Here, multi-fidelity Bayesian optimization (MFBO) \cite{poloczek2017multi,wu2020practical}, alternates between employing both a quick, low-fidelity model for \SI{3}{mm} resonators and a slower, high-fidelity model for \SI{3}{cm} ones. The high-fidelity predictions were eight times slower on average. 
We utilize finite element simulations with COMSOL~\cite{comsol} to maximize the quality factor. Despite the existence of analytical derivations for one-dimensional beams~\cite{fedorov2019generalized}, we discuss direct numerical simulations in this study. This choice aims to provide a more generalized design approach for large aspect-ratio resonators. The approach in this study could be expanded for other high Q resonators sensitive to the length scale but lacking derived analytical formulations~\cite{shinSpiderwebNanomechanicalResonators2022,bereyhiPerimeterModesNanomechanical2022,tsaturyanUltracoherentNanomechanicalResonators2017}.
The geometry of both resonators\cite{ghadimiElasticStrainEngineering2018} is parameterized (Supplementary Information Sec. 9) to allow the presence of a PnC with a defect embedded in the center and chosen to practically allow the most number of on-chip resonators. The method then quickly explores the design space by fast evaluations of the smaller structures at higher frequencies (MHz), while establishing a correlation with the large devices at lower frequencies. This enables it to probe (slow) solutions for the high-aspect-ratio structures only on rare occasions when it expects the design to achieve a large quality factor.

The effectiveness of MFBO relies on the correlation between low and high-fidelity models and their respective evaluation times. If the low-fidelity model lacks correlation with the high-fidelity model, or if its evaluation time is comparable, the method loses effectiveness. By parameterizing the design space independently from resonator length (see Methods), we observed a reasonable correlation between the \SI{3}{mm} (low-fidelity) and the \SI{3}{cm} (high-fidelity) resonator designs (Fig. \ref{fig:2_Design}a). 
The Q factor of the 3 cm design was a hundred times larger on average, 
confirming the expected correlation between the two models according to the soft clamped Q factor in \autoref{eq:Qtot}.
The variation in Figure \ref{fig:2_Design}b underscores that what improves performance for a \SI{3}{mm} resonator does not necessarily translate into a better \SI{3}{cm} resonator. Therefore, exclusively relying on the low fidelity model (\SI{3}{mm} resonator) would not yield the desired outcome for designing the \SI{3}{cm} resonator. Simultaneously, despite each resonator size having its unique set of optimal geometric parameters, the algorithm is still capable of learning enough from the response of the \SI{3}{mm} resonator to guide the optimization of the \SI{3}{cm} one.
Consequently, we applied MFBO to the log-normal Q factor, letting the algorithm selectively probe the design space via whichever fidelity it chooses. In essence, MFBO uses millimeter-scale simulations to guide centimeter-scale optimization.

Even though increasing the length of the resonator has a trend towards increasing the resonator's quality factor, optimization of the phononic crystal parameters is required for each given specific length scale.
The advantage of using multi-fidelity transformed Bayesian optimization is shown in Fig. \ref{fig:2_Design}d, as it finds a design with Q outperforming single-fidelity Bayesian optimization (regular BO) Q by approximately 20\%.
Figure \ref{fig:2_Design}c depicts the optimized geometry. After determining the optimal design, we focused on fabricating this high-aspect-ratio device.
The optimized geometry follows a tapering phononic crystal shape, which is similar to the result suggested by Ghadimi \textit{et al.} \cite{ghadimiElasticStrainEngineering2018}
The optimal design maximized the unit cell width around the clamping region and narrowed down the width of the unit cells coming near the center of the resonator.
One distinguishing aspect of our approach was that during the optimization under the design domain, the algorithm considered both symmetric and anti-symmetric modes, but the optimized mode was selected to be anti-symmetric rather than symmetric.
Because of this difference, the defect width was not minimized, unlike the early study \cite{ghadimiElasticStrainEngineering2018}, since the kinetic energy with significant movement is not maximized at the center.
All the designs considered during the optimization can be found in the Supplementary Video.
Once the best design was found, we focused on addressing the challenges of fabricating a device with such an extreme-aspect-ratio. 

\subsection*{Centimeter scale nanofabrication}

\begin{figure*} 
       \centering
	\includegraphics[width = 1\textwidth]{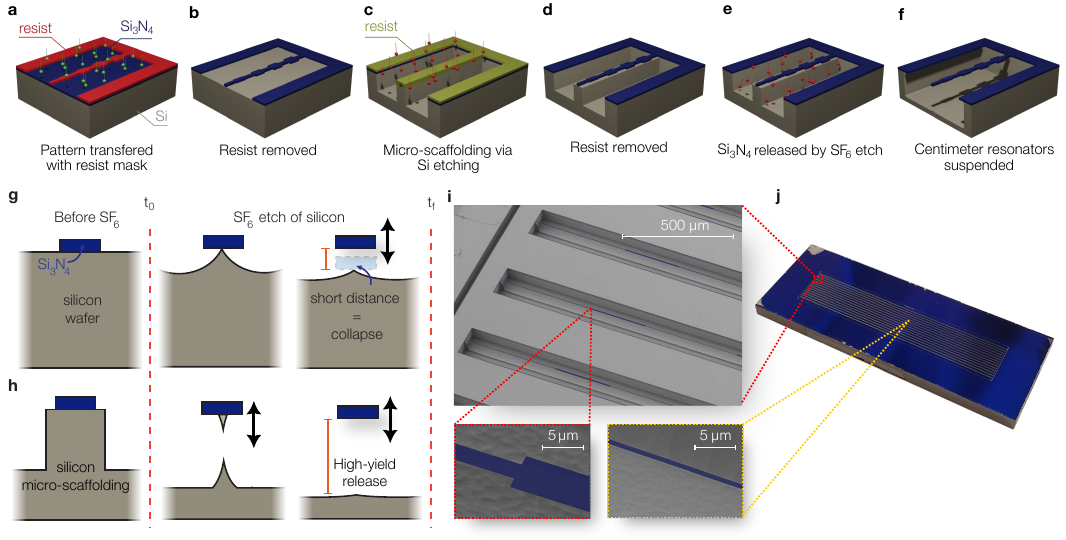} 
	\caption{\textbf{Fabrication of centimeter-scale nanoresonators}. \textbf{a-f} Schematic of the fabrication process composed of Si\textsubscript{3}N\textsubscript{4} patterning via dry etching \textbf{a}, mask removal \textbf{b},  cryogenic DRIE silicon etching with photoresist mask \textbf{c}, mask removal \textbf{d}, Si\textsubscript{3}N\textsubscript{4} undercut \textbf{e}, and Si\textsubscript{3}N\textsubscript{4} suspended over a large gap \textbf{f}. \textbf{g}, SF\textsubscript{6} release step when the Si\textsubscript{3}N\textsubscript{4} structure is suspended after being patterned, leading to collapse due to the short distance with the supporting silicon wafer. \textbf{h}, SF\textsubscript{6} release step with micro-scaffolding when the Si\textsubscript{3}N\textsubscript{4} structure is suspended after a DRIE of the supporting silicon wafer, resulting in a large gap and high-yield release. \textbf{i}, False-colored scanning electron microscope pictures of the optimized \SI{3}{cm} nanoresonators at the clamping area (top), at the boundary of the first unit cell (bottom-left), and at the center (bottom-right). The blue area indicates the area where Si\textsubscript{3}N\textsubscript{4} is suspended, consisting of the string and the overhang. \textbf{j}, Photograph of a chip containing fourteen  centimeter-scale nanoresonators, each \SI{3}{cm} long.}
	\label{fig:3_Fab}
\end{figure*}

Manufacturing centimeter-scale nanoresonators relies on fabrication intuition that shifts from conventional nanotechnologies in design principles, fabrication methods, and cost considerations. In accordance with Moore's Law, conventional nanotechnology has focused on miniaturization across all three dimensions (x, y, z). However, centimeter-scale nanotechnology marks a transition that requires components expanding out to macroscopic lengths in x and/or y while retaining their nanoscale thickness. These nanostructures not only have high-aspect-ratios and centimeters length at the macroscale, but they are also patterned at the nanoscale with small feature sizes, providing them with enhanced functionalities (mechanical, optical, etc.). In particular, our nanostrings are not just elongated and thin; they also incorporate precisely patterned phononic bandgaps.

In contrast to conventional miniaturization, the size constraints of centimeter-scale nanoresonators permit far fewer devices per wafer, requiring high fabrication yields due to the resulting higher costs per device. Given their long geometries, any fracture of a centimeter-scale nanoresonator not only results in fewer successful devices but these broken devices can also collapse over several neighboring structures, escalating the implications of fabrication errors and low yield. Moreover, the fabrication of these high-aspect-ratio structures requires them to be released with delicate nanofabrication techniques that do not exert any destructive forces during and after suspension. These processes must ensure that the high-aspect-ratio structures remain unfractured and undistorted, and are positioned safely away from nearby surfaces to prevent potential issues such as stiction due to attractive surface forces.

The schematic of the fabrication procedure is shown in Fig. \ref{fig:3_Fab}a-f (see Methods). First, high-stress (1.07GPa) Si\textsubscript{3}N\textsubscript{4} is deposited on a silicon wafer. A resist layer is spun on and lithographically patterned with an array of the \SI{3}{cm} long optimized PnC resonators. This is achieved by employing multiple exposures with varying resolutions to reduce the exposure time without compromising the accuracy (Supplementary Information Sec. 3). This patterned resist is used as a mask to transfer the design into the Si\textsubscript{3}N\textsubscript{4} film via directional plasma etch (Fig. \ref{fig:3_Fab}a). Typically the most crucial step is the careful release of these fragile structures by removing the silicon beneath the Si\textsubscript{3}N\textsubscript{4} resonator; with centimeter-scale nanostructures the requirements for successful suspension become much more stringent. 

A critical aspect to realizing these nanostructures is the high stress within Si\textsubscript{3}N\textsubscript{4} which not only contributes to achieving a high Q factor but also provides the required structural support and stability, allowing these taut strings to remain free-standing over remarkable distances without collapsing or sagging. In particular, we use dry SF\textsubscript{6} plasma etch to remove the silicon under the Si\textsubscript{3}N\textsubscript{4} resonator (Fig. \ref{fig:3_Fab}c-f) since it avoids the conventional stiction and collapse from surface tension present in liquid etchants\cite{norteNanofabricationOnChipOptical2015} and it doesn't leave any residues. Once released, these \SI{3}{cm}-long, \SI{70}{nm}-thick structures can displace tens of micrometers due to handling and static charge build up, and potentially collapse onto nearby surfaces. In combination with the SF\textsubscript{6} plasma dry release, we first engineer a micro-scaffolding\cite{ghadimiElasticStrainEngineering2018} (Fig. \ref{fig:3_Fab}g,h) into the silicon underneath our Si\textsubscript{3}N\textsubscript{4} that allows the free-standing structures to be suspended quickly, delicately, and far away from the substrate below, significantly increasing the yield and viability of this proposed nanotechnology (Supplementary Information Sec. 5). The micro-scaffolding is lithographically defined by a second exposure and transferred into the silicon substrate via deep directional plasma etch (Fig. \ref{fig:3_Fab}c). While our simulated strings are designed with a nominal thickness of \SI{50}{nm}, the fabricated resonators have a larger final thickness from edge to center along their length. This is caused by the difficulty of dissipating heat during the etching step of the silicon substrate, which is expected to decrease the Q factor predicted from the simulation (Supplementary Information Sec. 2).

Our choice to focus on 1D PnC nanostrings comes from a practical standpoint, which allows for packing numerous devices per chip  (Fig \ref{fig:3_Fab}i,j). By carefully engineering the delicate release of these structures, we achieved a fabrication yield as high as 93\% on our best chip and 75\% overall chips processed. While we only study 1D structures, the methodologies we developed are versatile and can be readily applied to more complex structures such as 2D phononic shields \cite{tsaturyanUltracoherentNanomechanicalResonators2017}.

\subsection*{Low dissipation at room temperature}

\begin{figure*} 
	\centering
	\includegraphics[width = 1\textwidth]{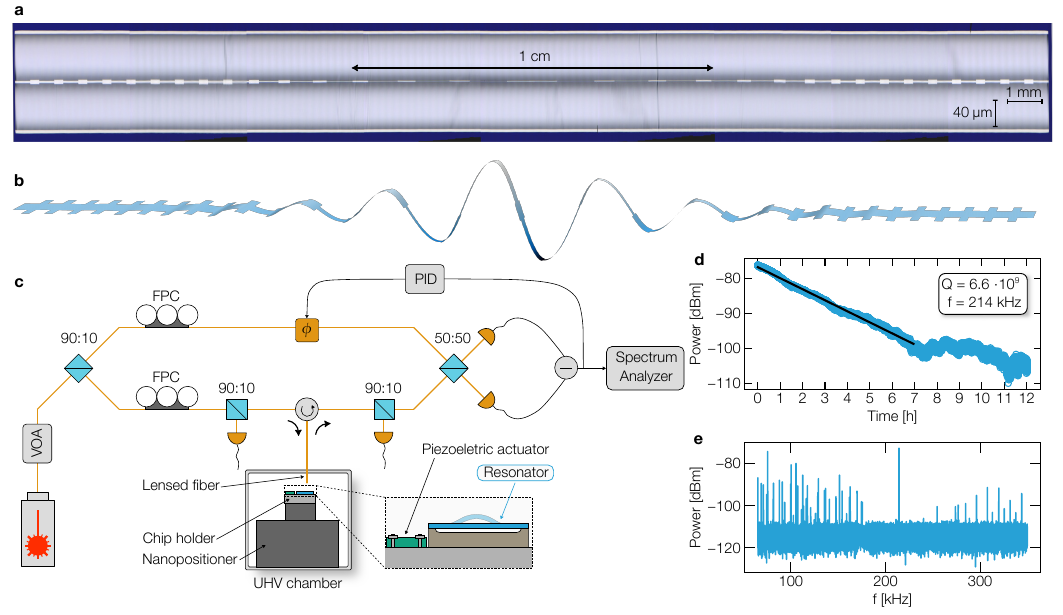} 
	\caption{\textbf{Quality factor validation}. \textbf{a}, Optical picture of the optimized \SI{3}{cm} nanoresonators. For illustration purposes, the picture is scaled along the vertical direction as indicated by the two scale bar. \textbf{b}, Mode shape of the eigenmode predicted to have high quality factor extracted from Finite Elements Analysis. \textbf{c}, Balanced homodyne optical interferometer built to experimentally measure the \SI{3}{cm} nanoresonators. The nanoresonator is placed inside an ultra-high vacuum (UHV) chamber and mechanically excited by a piezoelectric actuator. Its motion is detected by an infrared laser via a lensed fiber. VOA, variable optical attenuator. FPC, fiber polarization controller. PID, proportional integral derivative controller. $\Phi$, fiber stretcher. \textbf{d}, Ringdown trace of optimized nanoresonator excited at \SI{214}{kHz}. The black solid line indicates the linear fitting corresponding to an extracted quality factor of 6.6 billion at room temperature. \textbf{e}, Thermomechanical noise spectrum.}
	\label{fig:4_Exp}
\end{figure*}

To assess the mechanical properties of the fabricated nanoresonators, we characterized the strings’ out-of-plane displacements by a balanced homodyne optical interferometer (Fig. \ref{fig:4_Exp}c) built to experimentally measure the \SI{3}{cm} nanoresonators. The nanoresonator is placed inside an ultra-high vacuum (UHV) chamber at P $< 10^{-9}$ mbar to avoid gas damping, increasingly dominant for high-aspect-ratio structures (see Supplementary Information Sec. 1), and mechanically excited by a piezoelectric actuator. The resonator's motion is detected by an infrared laser via a lensed fiber. Figure \ref{fig:4_Exp}e shows the displacement spectrum obtained from a location near the center of the string, under thermal excitation, for the device depicted in Fig. \ref{fig:4_Exp}a. The spectrum shows a clear bandgap in frequency between \SI{175}{kHz} and \SI{255}{kHz} with one localized mode inside. The latter is observed at \SI{214}{kHz}, in good agreement with simulation prediction. Figure \ref{fig:4_Exp}b shows the predicted mode shapes obtained by finite element analysis for the eigenmode in the center of the bandgap. On the contrary, outside the \SI{175}{kHz} - \SI{255}{kHz} range, a plethora of modes are detected, whose displacement is distributed over the entire string length.

We experimentally evaluate the quality factor of the nanoresonators by applying a sinusoidal function to the piezoelectric actuator at a frequency near the eigenfrequency of the localized mode. Once the displacement at resonance reaches a plateau, the excitation is abruptly turned off to measure the ringdown of the mechanical mode. Figure \ref{fig:4_Exp}d shows the envelope of the obtained signal for our best performing device, where the measured decay rate is proportional to the nanoresonator energy dissipation and thus its quality factor. As Fig. \ref{fig:4_Exp}d shows, the localized mode at \SI{214}{kHz} decays for over 7 hours. This corresponds to a Q factor of 6.6 billion at room temperature. To corroborate these findings, we tested additional lithographically identical devices, exhibiting the same frequency response in good agreement with the simulation, and quality factors within a 50 \% range (Supplementary Information Sec. 10).

While our simulations expect Qs of 10 billion, the fabricated centimeter scale nanomechanical structures have high-aspect-ratios that make it difficult to dissipate heat during the (exothermic) undercut process (Fig \ref{fig:3_Fab}e), resulting in different Si\textsubscript{3}N\textsubscript{4} etch rates. This gives the nanostructures slightly different thickness and dimension from edge to center along the beam, thus reducing the fidelity between design and experiment with a lower measured Q (Supplementary Information Sec. 2).

\section*{DISCUSSION}

We demonstrated centimetre-scale nanomechanical resonators with aspect-ratio above $4.3 \times 10^5$. Our approach combines two features. First, using MFBO we are able to reduce the simulation cost while maintaining the required accuracy to precisely capture the resonator's behaviour. The resulting data-driven design process allows to optimize PnC strings obtaining soft-clamped modes which eliminate clamping losses and radiation to the substrate. Second, a dry etching technique that overcomes limitations such as stiction and collapse enables to reliably realize the optimized designs on-chip. The fabricated PnC strings extend for \SI{3}{cm} in length maintaining nanometers thickness and a minimum width of \SI{500}{nm}. With a Q of 6.6 billion at a frequency of \SI{214}{kHz}, we experimentally achieve the highest Q factor yet measured for clamped resonators at room temperature.

The obtained aspect-ratio enables not only to achieve a two-fold improvement of the quality factor in room-temperature environments, but it also leads to low resonance frequencies and large spacing between nearby mechanical modes. Those features translate into coherence time $t_{\rm coh} = \hbar Q/(k_{\rm B}T)$ approaching 1 ms and thermomechanical-limited force sensitivity of $\sqrt{4k_{\rm B}T\omega_{\rm m}m/Q} \sim$ aN Hz$^{-0.5}$.

A natural application is ground state cooling of the mechanical resonators in a room-temperature environment. The high coherence time enables these resonators to undergo more than $Q\omega_{\rm m}(\hbar/K_{\rm B}T) \approx$  200 coherent oscillations in the ground state before a thermal phonon enters the system \cite{aspelmeyerCavityOptomechanics2014}. The small thermal decoherence rate $\Gamma_{\rm th} = k_{\rm B}T/(\hbar Q)$ of a few Hz allows to resolve the zero point fluctuations $x_{\rm zp} = \sqrt{\hbar/(2m\omega_{\rm m})}$ with a displacement imprecision $S_{\rm xx}^{\rm imp} = 4x_{\rm zp}^2/\Gamma_{\rm th}$ below \SI{1e-15}{m} Hz$^{0.5}$, on pair with the imprecision limit due to shot noise achievable in conventional interferometer setup. This makes centimeter-scale nanoresonators particularly promising for the cavity-free cooling scheme\cite{plucharCavityfreeGroundstateCooling2020}. The developed resonators are also ideal candidates for creating high-precision sensors, specifically force detectors \cite{eichlerUltrahighQNanomechanicalResonators2022}, and hold promise for obtaining frequency stability on pair with state-of-the-art clocks \cite{sadeghiFrequencyFluctuationsNanomechanical2020, manzanequeResolutionLimitsResonant2023, lewisIntroductionFrequencyStandards1991}. Conservatively assuming sub-micrometers amplitude displacements in the linear regime, we can extrapolate a thermomechanical limited Allan deviation\cite{manzanequeResolutionLimitsResonant2023} $\sigma_y(\tau) \sim $ \num{3e-12}$/\sqrt{\tau}$ for $m_{\text{eff}} =$ \SI{4.96e-13}{Kg}.

Remarkably, the degree of acoustic isolation (quantified by the quality factor) we can achieve on a solid-state microchip is similar to values recently demonstrated for levitated nanoparticles operating at vacuum pressure levels more than two orders of magnitude lower than our resonators \cite{daniaUltrahighQualityFactor2023, magriniRealtimeOptimalQuantum2021}. This comparison gains further importance when considering that levitated particles are essentially isolated from the environment, interacting only minimally with residual gas molecules at vacuum levels as low as $10^{-11}$~mbar, which requires infrastructures different from the turbo and roughing pump combination used in this study (Supplementary Information Sec. 11). In constrast, our solid-state resonators are physically clamped to a room-temperature chip, surrounded by 100 times higher gas pressures and exhibit comparable acoustic dissipation.

The ability to combine macroscale resonators with nanomechanics offers unique possibilities to integrate the versatility of on-chip technology with the detection sensitivity of macroscale resonators. Notably, the only foreseeable limitation to producing even longer, higher-Q devices is that larger undercuts distance must be engineered, and practically going to longer, lower-frequency devices would require increasingly higher vacuum levels; this makes centimeter-scale nanotechnology particularly interesting for next-generation space applications\cite{el2020inertial,amaro2012low} which inherently operate at pressures below $10^{-9}$ mbar. Pushing the boundaries of fabrication capabilities with higher selectivity materials\cite{xu2023SiC,beccariStrainedCrystallineNanomechanical2022,manjeshwar2023InGap} would extend our current approach to more extreme-aspect-ratios and investigate unhackneyed physics. These include the exploration of weak forces such as ultralight dark matter \cite{manleySearchingVectorDark2021} and the investigation of gravitational effects at the nanoscale \cite{pratt_nanoscale_2023, matsumotoDemonstrationDisplacementSensing2019}. While Si\textsubscript{3}N\textsubscript{4} resonators have demonstrated the highest room-temperature quality factors, it is noteworthy that the quality factor for Si\textsubscript{3}N\textsubscript{4} resonators have been consistently reported to increase at cryogenic temperatures\cite{gisler2022soft,yuan2015silicon}. Based on conservative estimates, we predict that our centimeter-scale resonators could exhibit Q factors above \( 6 \times 10^{10} \) at cryogenic temperatures, potentially surpassing current cryogenic devices\cite{beccariStrainedCrystallineNanomechanical2022, maccabe2020nano}. By blurring the line between macroscopic and nanoscale objects, these centimeter-scale nanomechanical systems challenge our conventional intuitions about fabrication, costs, and computer design and promise to give us innovative capabilities which have not been available at smaller scales.

\section*{Methods}

\textbf{Computational experiments and design.} The design approach for high-aspect-ratio resonators was based on numerical analysis with COMSOL~\cite{comsol}. The quality factor was maximized via multi-fidelity Bayesian optimization without recurring to the analytical solution derived for beam-like Phononic Crystals (PnCs) \cite{fedorov2019generalized}.
In particular, we consider the trace-aware knowledge gradient (taKG) formulation of Bayesian optimization with two fidelities \cite{balandat2020botorch}. 
Detailed information about the formulation can be found in Supplementary Information Sec. 7.
The maximization based on the high-fidelity model becomes possible by learning the trend (surrogate model) from multiple low-fidelity predictions instead of using fewer high-fidelity evaluations. 
The approach is especially beneficial for cases when the difference in time evaluation between fidelities is significant, i.e., the time it takes to perform one function evaluation (one design prediction via COMSOL) for the high fidelity is much longer than for the lower fidelity. 

We considered a \SI{3}{cm} resonator as the high-fidelity model and a \SI{3}{mm} resonator as the low-fidelity model for the MFBO.
As mentioned in the main text, we expected them to be correlated, given that $Q \propto L^2$ (\autoref{eq:Qtot}) for the string type resonators neglecting the sharp curvature change around the clamping region using the PnC.
This correlation allows us to predict the response of the \SI{3}{cm} computationally expensive model by the \SI{3}{mm} relatively cheap model. 
For the centimeter-scale PnC resonator's quality factor maximization, we designed the resonator's geometry with a two-dimensional model. 
The model has nine design parameters, including five determining the resonator's overall shape, the unit cell's width and length ratio, and the defect's length and width. 
Design variables were set to be independent of the resonator length. 
Detailed parameter descriptions can be found in Supplementary Information Sec. 9.

Figure \ref{fig:2_Design}a shows the quality factor distribution obtained from randomly selected 25 high-fidelity PnC resonator designs and the same number for low-fidelity ones. 
For both lengths, the same design parameters are considered. 
Both length scale's quality factor follows a log-normal distribution. 
More importantly, the ratio between the two fidelities also follows a log-normal distribution as depicted in the histogram in Fig. \ref{fig:2_Design}b. 
The result indicates that the Q factor of the \SI{3}{cm} design is, on average a hundred times larger than that of the same design scaled down to \SI{3}{mm}, which confirms the expected correlation between the two models.
Nevertheless, the ratio shows significant variance, ensuring that the optimum design for the \SI{3}{mm} resonator does not precisely correspond to the best design for the \SI{3}{cm} case. 
These findings underscore that using MFBO with low- and high-fidelity simulation models leads to a balance between obtaining the required accuracy and minimizing the simulation cost. 

Figures \ref{fig:2_Design}d-f show the optimization iteration history considering that the high-fidelity simulation costs eight times more than the low-fidelity simulation. 
This average time difference between fidelities is determined from the initial designs obtained by random search. 
Figure \ref{fig:2_Design}d compares the results when the logarithm of the quality factor is considered for the maximization using multi-fidelity Bayesian optimization (transformed MFBO) and when single-fidelity Bayesian optimization is directly optimizing the quality factor (regular BO).
After starting with 25 randomly selected initial calculations for the \SI{3}{cm} and \SI{3}{mm} model (or only the \SI{3}{cm} model for the regular BO), the algorithm maximizes the quality factor for the \SI{3}{cm} model by searching for the best possible design parameters. We note that the initial random design of experiments affects the optimization performance, but in most cases, MFBO outperformed single-fidelity Bayesian optimization.
The results with different random initials comparing the transformed MFBO and regular BO are summarized in Supplementary Information Sec. 8.

Detailed information on the transformed MFBO is shown in Figs. \ref{fig:2_Design}e and f. 
Figure \ref{fig:2_Design}e illustrates the quality factor calculated for each fidelity, and Fig. \ref{fig:2_Design}f illustrates the distance from a previous optimized point to the point considered in that iteration. 
Right after the random search, the algorithm runs predominantly low-fidelity simulations to optimize the quality factor, taking advantage of the relatively cheap simulation cost. 
We note that the low-fidelity simulation has higher quality factor variance when compared to the high-fidelity simulations, given the larger number of designs being explored in the former vs. the latter.
The high values in the distance from the argmax plot (Fig. \ref{fig:2_Design}f) further confirm this by showing that the optimization is found not only by exploitation but also by exploration.
For example, the exploration phase improves the quality factor as observed at iteration $\times$ cost $\sim$ 800. \newline

\textbf{Nanofabrication for centimeter-scale PnC resonators.} The \SI{3}{cm} optimum design is fabricated on high stress silicon nitride (Si\textsubscript{3}N\textsubscript{4}), deposited by low-pressure chemical vapor deposition (LPCVD) on 2 mm silicon wafers. The fabrication starts by transferring the desired geometry on a thin positive tone resist (AR-P 6200) by a lithographic step. Typically, electron beam lithography or photolithography is used to pattern the masking layer. Electron beam lithography allows higher resolution and smaller feature sizes compared to the optical counterpart, but it is prone to stitching errors for structures exceeding the writing field. With each writing field extending for \SI{100}{\micro m} - \SI{1}{mm}, our centimeter-resonators require more than 30 fields, leading to noticeable stitching errors. We then implemented an overlap between adjacent writing fields of \SI{100}{nm} and controlled the dose at specific locations (more details can be found in Supplementary Information Sec. 3). This resulted in an accurate transfer of the desired geometry avoiding the presence of stitching errors. Despite the electron beam lithography superior resolution, photolithography is preferable for a fast and cost-effective manufacture at the large scale. With this in mind, we constrained the minimum feature size of the nanoresonators at \SI{500}{nm}, compatible with ultraviolet photolithography.

Next, the pattern is transferred to the Si\textsubscript{3}N\textsubscript{4} layer using an inductively coupled plasma (ICP) etching process (CHF\textsubscript{3} + O\textsubscript{2}) at room temperature (Fig. \ref{fig:3_Fab}a) before removing the masking layer (Fig. \ref{fig:3_Fab}b). 

The most critical part of the process is then suspending the high-aspect-ratio fragile structures over the substrate without causing any fracture, stiction or collapse. Typically this step is performed by liquid etchants such as KOH which selectively removes the silicon substrate. However, turbulences and surface-tension forces can lead to collapse destroying the suspended structures \cite{norteNanofabricationOnChipOptical2015}. Those forces depend on the surface area of the nanoresonators and thus increase with the aspect-ratio, drastically reducing the fabrication yield for centimeter-scale nanoresonators. To overcome those limitations, stiction-free dry release can be employed \cite{shinSpiderwebNanomechanicalResonators2022, mouraCentimeterscaleSuspendedPhotonic2018a}, where the silicon substrate is isotropically removed by plasma etching. Fluorine based (SF\textsubscript{6}) dry etching at cryogenic temperature is particularly suited in view of its high selectivity against Si\textsubscript{3}N\textsubscript{4}, for which it does not require any mask or additional cleaning steps. Nevertheless, geometries with extreme high-aspect-ratio require a large opening ($>$ 50 \textmu m) from the substrate to avoid attraction due to charging effects, not achievable by SF\textsubscript{6} plasma etching alone.

To this end, we first directionally etch the silicon substrate by employing a thick positive tone photoresist (S1813) as a proactive layer (Fig. \ref{fig:3_Fab}c). The step is carried out with cryogenic deep reactive ion etching (DRIE) using SF\textsubscript{6} + O\textsubscript{2} plasma \cite{maduroMolybdenumNanopillarArrays2021}. Cryogenic DRIE allows to control the opening size from the substrate without affecting the Si\textsubscript{3}N\textsubscript{4} film quality. However, photoresist is vulnerable to cracking in cryogenic DRIE \cite{sainiemiMaskMaterialEffects2007}. To circumvent this limitation, the centimeter scale nanoresonators are shielded by an outer ring, which stops the cracks and prevents them from reaching the nanoresonators (see Supplementary Information Sec. 4 for details). The photoresist is then stripped off (Fig. \ref{fig:3_Fab}d) and a hot piranha solution consisting of sulfuric acid and hydrogen peroxide is employed to remove residual contaminants on the surface. After that, hydrofluoric acid solution allows to remove oxides from the surface, which would otherwise prevent an even release of nanoresonators. Finally, the centimeter scale nanoresonators are suspended by a short 32 seconds fluorine-based (SF6) dry etching step (Fig. \ref{fig:3_Fab}e-f) performed at \SI{-120}{\celsius}. The process isotropically etches the silicon substrate employing a pressure of \SI{10}{mbar}, an ICP power of \SI{2000}{W} and a gas flow of \SI{500}{sccm}, while the RF power is set to \SI{0}{W}.

The developed process enables to achieve a fabrication yield as high as 93 \% for our best chip, while the average value among all the fabricated devices is 75 \%. The main limiting factor is ensuring a particle-free surface prior to the SF\textsubscript{6}.\newline

\textbf{UHV Lensed-fiber Optical Setup}. To assess the mechanical properties of the fabricated nanoresonators, we characterized the strings’ out-of-plane displacements by a laser interferometer (Fig. \ref{fig:4_Exp}c). In it, 10\% of the laser power of a \SI{1550}{nm} infrared laser is focused on the nanoresonator. The reflected signal is collected by a lensed fiber to interfere with the local oscillator (LO) signal consisting of the remaining 90\% of the infrared laser. The output signal, proportional to the resonator's displacement, is read out by an electronic spectrum analyzer after being converted by a balanced photodetector. The same output signal acts as an error function for the feedback loop employed to stabilize the phase of the setup against slow fluctuations caused by mechanical and thermal drift. The feedback loop is implemented with a PID controller, which adjusts the phase of the LO signal.

To avoid gas damping, increasingly dominant for high-aspect-ratio structures (see Supplementary Information Sec. 1 for details), the nanoresonator is placed inside an ultra high vacuum chamber capable of reaching P $<$ \SI{1e-9}{mbar}. The vacuum chamber is equipped with a 3-axis nanopositioner which allows to align the device with respect to the lensed fiber. 

The chip is placed near a piezoelectric actuator which can vibrate out-of-plane. The latter allows to mechanically excite specific resonance frequencies of the nanoresonators. We then experimentally evaluate the quality factor of the nanoresonators by applying a sinusoidal function to the piezoelectric actuator at a frequency near the eigenfrequency of the localized mode. Once the displacement at resonance reaches a plateau, the excitation is abruptly turned off to measure the ringdown of the mechanical mode.

The power of the infrared laser can be manually adjusted by a variable optical attenuator, enabling to vary the laser power incident on the nanoresonators inside the vacuum chamber. This feature allows to perform ringdown measurements at different laser powers, crucial to rule out any optothermal or optical effects of the incident laser signal on the measured Q factor (Supplementary Information Sec. 6).

\section*{Data availability}
The data supporting the findings of this study are available in a Zenodo database with the DOI identifier \href{https://zenodo.org/doi/10.5281/zenodo.10518818}{10.5281/zenodo.10518818}

\section*{Code availability}
All code associated with this study is available from the corresponding authors, M.A.B. and R.N., upon request.

\bibliography{bibliography}

\begin{thebibliography}{100}
\expandafter\ifx\csname url\endcsname\relax
  \def\url#1{\texttt{#1}}\fi
\expandafter\ifx\csname urlprefix\endcsname\relax\def\urlprefix{URL }\fi
\providecommand{\bibinfo}[2]{#2}
\providecommand{\eprint}[2][]{\url{#2}}

\bibitem{abramoviciLIGOLaserInterferometer1992}
\bibinfo{author}{Abramovici, A.} \emph{et~al.}
\newblock \bibinfo{title}{{{LIGO}}: {{The Laser Interferometer
  Gravitational-Wave Observatory}}}.
\newblock \emph{\bibinfo{journal}{Science}} \textbf{\bibinfo{volume}{256}},
  \bibinfo{pages}{325--333} (\bibinfo{year}{1992}).

\bibitem{gonzalezBrownianMotionMass1994}
\bibinfo{author}{Gonz{\'a}lez, G.~I.} \& \bibinfo{author}{Saulson, P.~R.}
\newblock \bibinfo{title}{Brownian motion of a mass suspended by an anelastic
  wire}.
\newblock \emph{\bibinfo{journal}{The Journal of the Acoustical Society of
  America}} \textbf{\bibinfo{volume}{96}}, \bibinfo{pages}{207--212}
  (\bibinfo{year}{1994}).

\bibitem{sidlesMagneticResonanceForce1995}
\bibinfo{author}{Sidles, J.~A.} \emph{et~al.}
\newblock \bibinfo{title}{Magnetic resonance force microscopy}.
\newblock \emph{\bibinfo{journal}{Reviews of Modern Physics}}
  \textbf{\bibinfo{volume}{67}}, \bibinfo{pages}{249--265}
  (\bibinfo{year}{1995}).

\bibitem{binnigAtomicForceMicroscope1986}
\bibinfo{author}{Binnig, G.}, \bibinfo{author}{Quate, C.~F.} \&
  \bibinfo{author}{Gerber, {\relax Ch}.}
\newblock \bibinfo{title}{Atomic {{Force Microscope}}}.
\newblock \emph{\bibinfo{journal}{Physical Review Letters}}
  \textbf{\bibinfo{volume}{56}}, \bibinfo{pages}{930--933}
  (\bibinfo{year}{1986}).

\bibitem{rossiVectorialScanningForce2017}
\bibinfo{author}{Rossi, N.} \emph{et~al.}
\newblock \bibinfo{title}{Vectorial scanning force microscopy using a nanowire
  sensor}.
\newblock \emph{\bibinfo{journal}{Nature Nanotechnology}}
  \textbf{\bibinfo{volume}{12}}, \bibinfo{pages}{150--155}
  (\bibinfo{year}{2017}).

\bibitem{nicholNanomechanicalDetectionNuclear2012}
\bibinfo{author}{Nichol, J.~M.}, \bibinfo{author}{Hemesath, E.~R.},
  \bibinfo{author}{Lauhon, L.~J.} \& \bibinfo{author}{Budakian, R.}
\newblock \bibinfo{title}{Nanomechanical detection of nuclear magnetic
  resonance using a silicon nanowire oscillator}.
\newblock \emph{\bibinfo{journal}{Physical Review B}}
  \textbf{\bibinfo{volume}{85}}, \bibinfo{pages}{054414}
  (\bibinfo{year}{2012}).

\bibitem{harrisQuantumMechanicalSystems2015}
\bibinfo{author}{Harris, J.}, \bibinfo{author}{Rabl, P.} \&
  \bibinfo{author}{Schliesser, A.}
\newblock \bibinfo{title}{``{{Quantum}}'' {{Mechanical Systems}}: Bridging
  foundations and applications}.
\newblock \emph{\bibinfo{journal}{Annalen der Physik}}
  \textbf{\bibinfo{volume}{527}}, \bibinfo{pages}{A13--A14}
  (\bibinfo{year}{2015}).

\bibitem{guoFeedbackCoolingRoom2019}
\bibinfo{author}{Guo, J.}, \bibinfo{author}{Norte, R.} \&
  \bibinfo{author}{Gr{\"o}blacher, S.}
\newblock \bibinfo{title}{Feedback {{Cooling}} of a {{Room Temperature
  Mechanical Oscillator}} close to its {{Motional Ground State}}}.
\newblock \emph{\bibinfo{journal}{Physical Review Letters}}
  \textbf{\bibinfo{volume}{123}}, \bibinfo{pages}{223602}
  (\bibinfo{year}{2019}).

\bibitem{masonContinuousForceDisplacement2019}
\bibinfo{author}{Mason, D.}, \bibinfo{author}{Chen, J.},
  \bibinfo{author}{Rossi, M.}, \bibinfo{author}{Tsaturyan, Y.} \&
  \bibinfo{author}{Schliesser, A.}
\newblock \bibinfo{title}{Continuous force and displacement measurement below
  the standard quantum limit}.
\newblock \emph{\bibinfo{journal}{Nature Physics}}
  \textbf{\bibinfo{volume}{15}}, \bibinfo{pages}{745--749}
  (\bibinfo{year}{2019}).

\bibitem{saarinenLaserCoolingMembraneinthemiddle2023}
\bibinfo{author}{Saarinen, S.~A.}, \bibinfo{author}{Kralj, N.},
  \bibinfo{author}{Langman, E.~C.}, \bibinfo{author}{Tsaturyan, Y.} \&
  \bibinfo{author}{Schliesser, A.}
\newblock \bibinfo{title}{Laser cooling a membrane-in-the-middle system close
  to the quantum ground state from room temperature}.
\newblock \emph{\bibinfo{journal}{Optica}} \textbf{\bibinfo{volume}{10}},
  \bibinfo{pages}{364--372} (\bibinfo{year}{2023}).

\bibitem{seisGroundStateCooling2022a}
\bibinfo{author}{Seis, Y.} \emph{et~al.}
\newblock \bibinfo{title}{Ground state cooling of an ultracoherent
  electromechanical system}.
\newblock \emph{\bibinfo{journal}{Nature Communications}}
  \textbf{\bibinfo{volume}{13}}, \bibinfo{pages}{1507} (\bibinfo{year}{2022}).

\bibitem{midoloNanooptoelectromechanicalSystems2018}
\bibinfo{author}{Midolo, L.}, \bibinfo{author}{Schliesser, A.} \&
  \bibinfo{author}{Fiore, A.}
\newblock \bibinfo{title}{Nano-opto-electro-mechanical systems}.
\newblock \emph{\bibinfo{journal}{Nature Nanotechnology}}
  \textbf{\bibinfo{volume}{13}}, \bibinfo{pages}{11--18}
  (\bibinfo{year}{2018}).

\bibitem{magriniRealtimeOptimalQuantum2021}
\bibinfo{author}{Magrini, L.} \emph{et~al.}
\newblock \bibinfo{title}{Real-time optimal quantum control of mechanical
  motion at room temperature}.
\newblock \emph{\bibinfo{journal}{Nature}} \textbf{\bibinfo{volume}{595}},
  \bibinfo{pages}{373--377} (\bibinfo{year}{2021}).

\bibitem{manzanequeMethodDetermineClosedLoop2020}
\bibinfo{author}{Manzaneque, T.}, \bibinfo{author}{Steeneken, P.~G.},
  \bibinfo{author}{Alijani, F.} \& \bibinfo{author}{Ghatkesar, M.~K.}
\newblock \bibinfo{title}{Method to determine the closed-loop precision of
  resonant sensors from open-loop measurements}.
\newblock \emph{\bibinfo{journal}{IEEE Sensors Journal}}
  \textbf{\bibinfo{volume}{20}}, \bibinfo{pages}{14262--14272}
  (\bibinfo{year}{2020}).

\bibitem{chenDiamagneticallyLevitatingResonant2021}
\bibinfo{author}{Chen, X.}, \bibinfo{author}{Kothari, N.},
  \bibinfo{author}{Ke{\c s}kekler, A.}, \bibinfo{author}{Steeneken, P.~G.} \&
  \bibinfo{author}{Alijani, F.}
\newblock \bibinfo{title}{Diamagnetically levitating resonant weighing scale}.
\newblock \emph{\bibinfo{journal}{Sensors and Actuators A: Physical}}
  \textbf{\bibinfo{volume}{330}}, \bibinfo{pages}{112842}
  (\bibinfo{year}{2021}).

\bibitem{hanayInertialImagingNanomechanical2015}
\bibinfo{author}{Hanay, M.~S.} \emph{et~al.}
\newblock \bibinfo{title}{Inertial imaging with nanomechanical systems}.
\newblock \emph{\bibinfo{journal}{Nature Nanotechnology}}
  \textbf{\bibinfo{volume}{10}}, \bibinfo{pages}{339--344}
  (\bibinfo{year}{2015}).

\bibitem{reinhardt2016ultralow}
\bibinfo{author}{Reinhardt, C.}, \bibinfo{author}{M{\"u}ller, T.},
  \bibinfo{author}{Bourassa, A.} \& \bibinfo{author}{Sankey, J.~C.}
\newblock \bibinfo{title}{Ultralow-noise sin trampoline resonators for sensing
  and optomechanics}.
\newblock \emph{\bibinfo{journal}{Physical Review X}}
  \textbf{\bibinfo{volume}{6}}, \bibinfo{pages}{021001} (\bibinfo{year}{2016}).

\bibitem{eichlerUltrahighQNanomechanicalResonators2022}
\bibinfo{author}{Eichler, A.}
\newblock \bibinfo{title}{Ultra-high-{{Q}} nanomechanical resonators for force
  sensing}.
\newblock \emph{\bibinfo{journal}{Materials for Quantum Technology}}
  \textbf{\bibinfo{volume}{2}}, \bibinfo{pages}{043001} (\bibinfo{year}{2022}).

\bibitem{cagnoliDampingDilutionFactor2000}
\bibinfo{author}{Cagnoli, G.} \emph{et~al.}
\newblock \bibinfo{title}{Damping dilution factor for a pendulum in an
  interferometric gravitational waves detector}.
\newblock \emph{\bibinfo{journal}{Physics Letters A}}
  \textbf{\bibinfo{volume}{272}}, \bibinfo{pages}{39--45}
  (\bibinfo{year}{2000}).

\bibitem{unterreithmeierDampingNanomechanicalResonators2010}
\bibinfo{author}{Unterreithmeier, Q.~P.}, \bibinfo{author}{Faust, T.} \&
  \bibinfo{author}{Kotthaus, J.~P.}
\newblock \bibinfo{title}{Damping of {{Nanomechanical Resonators}}}.
\newblock \emph{\bibinfo{journal}{Physical Review Letters}}
  \textbf{\bibinfo{volume}{105}}, \bibinfo{pages}{027205}
  (\bibinfo{year}{2010}).

\bibitem{schmidDampingMechanismsHighQ2011}
\bibinfo{author}{Schmid, S.}, \bibinfo{author}{Jensen, K.},
  \bibinfo{author}{K~Nielsen, {\relax KH}.} \& \bibinfo{author}{Boisen, A.}
\newblock \bibinfo{title}{Damping mechanisms in high-{{Q}} micro and
  nanomechanical string resonators}.
\newblock \emph{\bibinfo{journal}{Physical Review B}}
  \textbf{\bibinfo{volume}{84}}, \bibinfo{pages}{1--6} (\bibinfo{year}{2011}).

\bibitem{yuControlMaterialDamping2012}
\bibinfo{author}{Yu, P.-L.}, \bibinfo{author}{Purdy, T.~P.} \&
  \bibinfo{author}{Regal, C.~A.}
\newblock \bibinfo{title}{Control of {{Material Damping}} in {{High- Q Membrane
  Microresonators}}}.
\newblock \emph{\bibinfo{journal}{Physical Review Letters}}
  \textbf{\bibinfo{volume}{108}}, \bibinfo{pages}{083603}
  (\bibinfo{year}{2012}).

\bibitem{cummingDesignDevelopmentAdvanced2012}
\bibinfo{author}{Cumming, A.~V.} \emph{et~al.}
\newblock \bibinfo{title}{Design and development of the advanced {{LIGO}}
  monolithic fused silica suspension}.
\newblock \emph{\bibinfo{journal}{Classical and Quantum Gravity}}
  \textbf{\bibinfo{volume}{29}}, \bibinfo{pages}{035003}
  (\bibinfo{year}{2012}).

\bibitem{dawidInvestigationViolinMode1997}
\bibinfo{author}{Dawid, D.~J.} \& \bibinfo{author}{Kawamura, S.}
\newblock \bibinfo{title}{Investigation of violin mode {{Q}} for wires of
  various materials}.
\newblock \emph{\bibinfo{journal}{Review of Scientific Instruments}}
  \textbf{\bibinfo{volume}{68}}, \bibinfo{pages}{4600--4603}
  (\bibinfo{year}{1997}).

\bibitem{whittleApproachingMotionalGround2021a}
\bibinfo{author}{Whittle, C.} \emph{et~al.}
\newblock \bibinfo{title}{Approaching the motional ground state of a 10-kg
  object}.
\newblock \emph{\bibinfo{journal}{Science}} \textbf{\bibinfo{volume}{372}},
  \bibinfo{pages}{1333--1336} (\bibinfo{year}{2021}).

\bibitem{abbott2016observation}
\bibinfo{author}{Abbott, B.~P.} \emph{et~al.}
\newblock \bibinfo{title}{Observation of gravitational waves from a binary
  black hole merger}.
\newblock \emph{\bibinfo{journal}{Physical review letters}}
  \textbf{\bibinfo{volume}{116}}, \bibinfo{pages}{061102}
  (\bibinfo{year}{2016}).

\bibitem{corbittOpticalDilutionFeedback2007}
\bibinfo{author}{Corbitt, T.} \emph{et~al.}
\newblock \bibinfo{title}{Optical {{Dilution}} and {{Feedback Cooling}} of a
  {{Gram-Scale Oscillator}} to 6.9 {{mK}}}.
\newblock \emph{\bibinfo{journal}{Physical Review Letters}}
  \textbf{\bibinfo{volume}{99}}, \bibinfo{pages}{160801}
  (\bibinfo{year}{2007}).

\bibitem{matsumotoDemonstrationDisplacementSensing2019}
\bibinfo{author}{Matsumoto, N.} \emph{et~al.}
\newblock \bibinfo{title}{Demonstration of displacement sensing of a mg-scale
  pendulum for mm- and mg- scale gravity measurements}.
\newblock \emph{\bibinfo{journal}{Physical Review Letters}}
  \textbf{\bibinfo{volume}{122}}, \bibinfo{pages}{071101}
  (\bibinfo{year}{2019}).
\newblock \eprint{1809.05081}.

\bibitem{westerveld2021sensitive}
\bibinfo{author}{Westerveld, W.~J.} \emph{et~al.}
\newblock \bibinfo{title}{Sensitive, small, broadband and scalable
  optomechanical ultrasound sensor in silicon photonics}.
\newblock \emph{\bibinfo{journal}{Nature Photonics}}
  \textbf{\bibinfo{volume}{15}}, \bibinfo{pages}{341--345}
  (\bibinfo{year}{2021}).

\bibitem{norteMechanicalResonatorsQuantum2016}
\bibinfo{author}{Norte, R.~A.}, \bibinfo{author}{Moura, J.~P.} \&
  \bibinfo{author}{Gr\"oblacher, S.}
\newblock \bibinfo{title}{Mechanical resonators for quantum optomechanics
  experiments at room temperature}.
\newblock \emph{\bibinfo{journal}{Phys. Rev. Lett.}}
  \textbf{\bibinfo{volume}{116}}, \bibinfo{pages}{147202}
  (\bibinfo{year}{2016}).

\bibitem{bereyhiHierarchicalTensileStructures2022}
\bibinfo{author}{Bereyhi, M.~J.} \emph{et~al.}
\newblock \bibinfo{title}{Hierarchical tensile structures with ultralow
  mechanical dissipation}.
\newblock \emph{\bibinfo{journal}{Nature Communications}}
  \textbf{\bibinfo{volume}{13}}, \bibinfo{pages}{3097} (\bibinfo{year}{2022}).

\bibitem{prattNanoscaleTorsionalDissipation2023}
\bibinfo{author}{Pratt, J.~R.} \emph{et~al.}
\newblock \bibinfo{title}{Nanoscale {{Torsional Dissipation Dilution}} for
  {{Quantum Experiments}} and {{Precision Measurement}}}.
\newblock \emph{\bibinfo{journal}{Physical Review X}}
  \textbf{\bibinfo{volume}{13}}, \bibinfo{pages}{011018}
  (\bibinfo{year}{2023}).

\bibitem{bereyhiPerimeterModesNanomechanical2022}
\bibinfo{author}{Bereyhi, M.~J.} \emph{et~al.}
\newblock \bibinfo{title}{Perimeter {{Modes}} of {{Nanomechanical Resonators
  Exhibit Quality Factors Exceeding}} \$\{10\}\^\{9\}\$ at {{Room
  Temperature}}}.
\newblock \emph{\bibinfo{journal}{Physical Review X}}
  \textbf{\bibinfo{volume}{12}}, \bibinfo{pages}{021036}
  (\bibinfo{year}{2022}).

\bibitem{shinSpiderwebNanomechanicalResonators2022}
\bibinfo{author}{Shin, D.} \emph{et~al.}
\newblock \bibinfo{title}{Spiderweb {{Nanomechanical Resonators}} via
  {{Bayesian Optimization}}: {{Inspired}} by {{Nature}} and {{Guided}} by
  {{Machine Learning}}}.
\newblock \emph{\bibinfo{journal}{Advanced Materials}}
  \textbf{\bibinfo{volume}{34}}, \bibinfo{pages}{2106248}
  (\bibinfo{year}{2022}).

\bibitem{tsaturyanUltracoherentNanomechanicalResonators2017}
\bibinfo{author}{Tsaturyan, Y.}, \bibinfo{author}{Barg, A.},
  \bibinfo{author}{Polzik, E.~S.} \& \bibinfo{author}{Schliesser, A.}
\newblock \bibinfo{title}{Ultracoherent nanomechanical resonators via soft
  clamping and dissipation dilution}.
\newblock \emph{\bibinfo{journal}{Nature Nanotechnology}}
  \textbf{\bibinfo{volume}{12}}, \bibinfo{pages}{776--783}
  (\bibinfo{year}{2017}).

\bibitem{ghadimiElasticStrainEngineering2018}
\bibinfo{author}{Ghadimi, A.~H.} \emph{et~al.}
\newblock \bibinfo{title}{Elastic strain engineering for ultralow mechanical
  dissipation}.
\newblock \emph{\bibinfo{journal}{Science}} \textbf{\bibinfo{volume}{360}},
  \bibinfo{pages}{764--768} (\bibinfo{year}{2018}).

\bibitem{rossiMeasurementbasedQuantumControl2018}
\bibinfo{author}{Rossi, M.}, \bibinfo{author}{Mason, D.},
  \bibinfo{author}{Chen, J.}, \bibinfo{author}{Tsaturyan, Y.} \&
  \bibinfo{author}{Schliesser, A.}
\newblock \bibinfo{title}{Measurement-based quantum control of mechanical
  motion}.
\newblock \emph{\bibinfo{journal}{Nature}} \textbf{\bibinfo{volume}{563}},
  \bibinfo{pages}{53--58} (\bibinfo{year}{2018}).

\bibitem{purdyObservationRadiationPressure2013}
\bibinfo{author}{Purdy, T.~P.}, \bibinfo{author}{Peterson, R.~W.} \&
  \bibinfo{author}{Regal, C.~A.}
\newblock \bibinfo{title}{Observation of {{Radiation Pressure Shot Noise}} on a
  {{Macroscopic Object}}}.
\newblock \emph{\bibinfo{journal}{Science}} \textbf{\bibinfo{volume}{339}},
  \bibinfo{pages}{801--804} (\bibinfo{year}{2013}).

\bibitem{halgMembraneBasedScanningForce2021}
\bibinfo{author}{H{\"a}lg, D.} \emph{et~al.}
\newblock \bibinfo{title}{Membrane-{{Based Scanning Force Microscopy}}}.
\newblock \emph{\bibinfo{journal}{Physical Review Applied}}
  \textbf{\bibinfo{volume}{15}}, \bibinfo{pages}{L021001}
  (\bibinfo{year}{2021}).

\bibitem{chasteNanomechanicalMassSensor2012}
\bibinfo{author}{Chaste, J.} \emph{et~al.}
\newblock \bibinfo{title}{A nanomechanical mass sensor with yoctogram
  resolution}.
\newblock \emph{\bibinfo{journal}{Nature Nanotechnology}}
  \textbf{\bibinfo{volume}{7}}, \bibinfo{pages}{301--304}
  (\bibinfo{year}{2012}).

\bibitem{krause2012high}
\bibinfo{author}{Krause, A.~G.}, \bibinfo{author}{Winger, M.},
  \bibinfo{author}{Blasius, T.~D.}, \bibinfo{author}{Lin, Q.} \&
  \bibinfo{author}{Painter, O.}
\newblock \bibinfo{title}{A high-resolution microchip optomechanical
  accelerometer}.
\newblock \emph{\bibinfo{journal}{Nature Photonics}}
  \textbf{\bibinfo{volume}{6}}, \bibinfo{pages}{768--772}
  (\bibinfo{year}{2012}).

\bibitem{naserbakht2019squeeze}
\bibinfo{author}{Naserbakht, S.} \& \bibinfo{author}{Dantan, A.}
\newblock \bibinfo{title}{Squeeze film pressure sensors based on sin membrane
  sandwiches}.
\newblock \emph{\bibinfo{journal}{Sensors and Actuators A: Physical}}
  \textbf{\bibinfo{volume}{298}}, \bibinfo{pages}{111588}
  (\bibinfo{year}{2019}).

\bibitem{scherschligt2018quantum}
\bibinfo{author}{Scherschligt, J.} \emph{et~al.}
\newblock \bibinfo{title}{Quantum-based vacuum metrology at the national
  institute of standards and technology}.
\newblock \emph{\bibinfo{journal}{Journal of Vacuum Science \& Technology A}}
  \textbf{\bibinfo{volume}{36}}, \bibinfo{pages}{040801}
  (\bibinfo{year}{2018}).

\bibitem{huang2023room}
\bibinfo{author}{Huang, G.}, \bibinfo{author}{Beccari, A.},
  \bibinfo{author}{Engelsen, N.~J.} \& \bibinfo{author}{Kippenberg, T.~J.}
\newblock \bibinfo{title}{Room-temperature quantum optomechanics using an
  ultralow noise cavity}.
\newblock \emph{\bibinfo{journal}{Nature}} \textbf{\bibinfo{volume}{626}},
  \bibinfo{pages}{512--516} (\bibinfo{year}{2024}).

\bibitem{purdyQuantumCorrelationsRoomtemperature2017}
\bibinfo{author}{Purdy, T.~P.}, \bibinfo{author}{Grutter, K.~E.},
  \bibinfo{author}{Srinivasan, K.} \& \bibinfo{author}{Taylor, J.~M.}
\newblock \bibinfo{title}{Quantum correlations from a room-temperature
  optomechanical cavity}.
\newblock \emph{\bibinfo{journal}{Science}} \textbf{\bibinfo{volume}{356}},
  \bibinfo{pages}{1265--1268} (\bibinfo{year}{2017}).

\bibitem{kristensenLonglivedEfficientOptomechanical2023}
\bibinfo{author}{Kristensen, M.~B.}, \bibinfo{author}{Kralj, N.},
  \bibinfo{author}{Langman, E.~C.} \& \bibinfo{author}{Schliesser, A.}
\newblock \bibinfo{title}{Long-lived and efficient optomechanical memory for
  light}.
\newblock \emph{\bibinfo{journal}{Physical Review Letters}}
  \textbf{\bibinfo{volume}{132}}, \bibinfo{pages}{100802}
  (\bibinfo{year}{2024}).

\bibitem{zhangZeptonewtonForceSensing2022}
\bibinfo{author}{Zhang, S.-D.} \emph{et~al.}
\newblock \bibinfo{title}{Zeptonewton force sensing with squeezed quadratic
  optomechanics}.
\newblock \emph{\bibinfo{journal}{arXiv preprint arXiv:2202.08690}}
  (\bibinfo{year}{2022}).

\bibitem{manleySearchingVectorDark2021}
\bibinfo{author}{Manley, J.}, \bibinfo{author}{Chowdhury, M.~D.},
  \bibinfo{author}{Grin, D.}, \bibinfo{author}{Singh, S.} \&
  \bibinfo{author}{Wilson, D.~J.}
\newblock \bibinfo{title}{Searching for {{Vector Dark Matter}} with an
  {{Optomechanical Accelerometer}}}.
\newblock \emph{\bibinfo{journal}{Physical Review Letters}}
  \textbf{\bibinfo{volume}{126}}, \bibinfo{pages}{061301}
  (\bibinfo{year}{2021}).

\bibitem{pate2020casimir}
\bibinfo{author}{Pate, J.~M.}, \bibinfo{author}{Goryachev, M.},
  \bibinfo{author}{Chiao, R.~Y.}, \bibinfo{author}{Sharping, J.~E.} \&
  \bibinfo{author}{Tobar, M.~E.}
\newblock \bibinfo{title}{Casimir spring and dilution in macroscopic cavity
  optomechanics}.
\newblock \emph{\bibinfo{journal}{Nature Physics}}
  \textbf{\bibinfo{volume}{16}}, \bibinfo{pages}{1117--1122}
  (\bibinfo{year}{2020}).

\bibitem{fong2019phonon}
\bibinfo{author}{Fong, K.~Y.} \emph{et~al.}
\newblock \bibinfo{title}{Phonon heat transfer across a vacuum through quantum
  fluctuations}.
\newblock \emph{\bibinfo{journal}{Nature}} \textbf{\bibinfo{volume}{576}},
  \bibinfo{pages}{243--247} (\bibinfo{year}{2019}).

\bibitem{pearson2021measuring}
\bibinfo{author}{Pearson, A.~N.} \emph{et~al.}
\newblock \bibinfo{title}{Measuring the thermodynamic cost of timekeeping}.
\newblock \emph{\bibinfo{journal}{Physical Review X}}
  \textbf{\bibinfo{volume}{11}}, \bibinfo{pages}{021029}
  (\bibinfo{year}{2021}).

\bibitem{daniaUltrahighQualityFactor2023}
\bibinfo{author}{Dania, L.} \emph{et~al.}
\newblock \bibinfo{title}{Ultrahigh quality factor of a levitated
  nanomechanical oscillator}.
\newblock \emph{\bibinfo{journal}{Physical Review Letters}}
  \textbf{\bibinfo{volume}{132}}, \bibinfo{pages}{133602}
  (\bibinfo{year}{2024}).

\bibitem{beccariStrainedCrystallineNanomechanical2022}
\bibinfo{author}{Beccari, A.} \emph{et~al.}
\newblock \bibinfo{title}{Strained crystalline nanomechanical resonators with
  quality factors above 10 billion}.
\newblock \emph{\bibinfo{journal}{Nature Physics}}
  \textbf{\bibinfo{volume}{18}}, \bibinfo{pages}{436--441}
  (\bibinfo{year}{2022}).

\bibitem{maccabe2020nano}
\bibinfo{author}{MacCabe, G.~S.} \emph{et~al.}
\newblock \bibinfo{title}{Nano-acoustic resonator with ultralong phonon
  lifetime}.
\newblock \emph{\bibinfo{journal}{Science}} \textbf{\bibinfo{volume}{370}},
  \bibinfo{pages}{840--843} (\bibinfo{year}{2020}).

\bibitem{de2022mechanical}
\bibinfo{author}{de~Jong, M.~H.} \emph{et~al.}
\newblock \bibinfo{title}{Mechanical dissipation by substrate--mode coupling in
  sin resonators}.
\newblock \emph{\bibinfo{journal}{Applied Physics Letters}}
  \textbf{\bibinfo{volume}{121}}, \bibinfo{pages}{032201}
  (\bibinfo{year}{2022}).

\bibitem{fedorov2019generalized}
\bibinfo{author}{Fedorov, S.~A.} \emph{et~al.}
\newblock \bibinfo{title}{Generalized dissipation dilution in strained
  mechanical resonators}.
\newblock \emph{\bibinfo{journal}{Physical Review B}}
  \textbf{\bibinfo{volume}{99}}, \bibinfo{pages}{054107}
  (\bibinfo{year}{2019}).

\bibitem{thompson2008strong}
\bibinfo{author}{Thompson, J.} \emph{et~al.}
\newblock \bibinfo{title}{Strong dispersive coupling of a high-finesse cavity
  to a micromechanical membrane}.
\newblock \emph{\bibinfo{journal}{Nature}} \textbf{\bibinfo{volume}{452}},
  \bibinfo{pages}{72--75} (\bibinfo{year}{2008}).

\bibitem{poloczek2017multi}
\bibinfo{author}{Poloczek, M.}, \bibinfo{author}{Wang, J.} \&
  \bibinfo{author}{Frazier, P.}
\newblock \bibinfo{title}{Multi-information source optimization}.
\newblock In \bibinfo{editor}{Guyon, I.} \emph{et~al.} (eds.)
  \emph{\bibinfo{booktitle}{Advances in Neural Information Processing
  Systems}}, vol.~\bibinfo{volume}{30} (\bibinfo{publisher}{Curran Associates,
  Inc.}, \bibinfo{year}{2017}).

\bibitem{wu2020practical}
\bibinfo{author}{Wu, J.}, \bibinfo{author}{Toscano-Palmerin, S.},
  \bibinfo{author}{Frazier, P.~I.} \& \bibinfo{author}{Wilson, A.~G.}
\newblock \bibinfo{title}{Practical multi-fidelity bayesian optimization for
  hyperparameter tuning}.
\newblock In \bibinfo{editor}{Adams, R.~P.} \& \bibinfo{editor}{Gogate, V.}
  (eds.) \emph{\bibinfo{booktitle}{Proceedings of The 35th Uncertainty in
  Artificial Intelligence Conference}}, vol. \bibinfo{volume}{115} of
  \emph{\bibinfo{series}{Proceedings of Machine Learning Research}},
  \bibinfo{pages}{788--798} (\bibinfo{publisher}{PMLR}, \bibinfo{year}{2020}).

\bibitem{comsol}
\bibinfo{author}{Inc., C.}
\newblock \bibinfo{title}{Comsol} (\bibinfo{year}{2020}).
\newblock \urlprefix\url{http://www.comsol.com/products/multiphysics/}.

\bibitem{norteNanofabricationOnChipOptical2015}
\bibinfo{author}{Norte, R.~A.}
\newblock \emph{\bibinfo{title}{Nanofabrication for {{On-Chip Optical
  Levitation}}, {{Atom-Trapping}}, and {{Superconducting Quantum Circuits}}}}.
\newblock Ph.D. thesis, \bibinfo{school}{California Institute of Technology}
  (\bibinfo{year}{2015}).

\bibitem{aspelmeyerCavityOptomechanics2014}
\bibinfo{author}{Aspelmeyer, M.}, \bibinfo{author}{Kippenberg, T.~J.} \&
  \bibinfo{author}{Marquardt, F.}
\newblock \bibinfo{title}{Cavity optomechanics}.
\newblock \emph{\bibinfo{journal}{Reviews of Modern Physics}}
  \textbf{\bibinfo{volume}{86}}, \bibinfo{pages}{1391--1452}
  (\bibinfo{year}{2014}).

\bibitem{plucharCavityfreeGroundstateCooling2020}
\bibinfo{author}{Pluchar, C.~M.}, \bibinfo{author}{Agrawal, A.~R.},
  \bibinfo{author}{Schenk, E.}, \bibinfo{author}{Wilson, D.~J.} \&
  \bibinfo{author}{Wilson, D.~J.}
\newblock \bibinfo{title}{Towards cavity-free ground-state cooling of an
  acoustic-frequency silicon nitride membrane}.
\newblock \emph{\bibinfo{journal}{Applied Optics}}
  \textbf{\bibinfo{volume}{59}}, \bibinfo{pages}{G107--G111}
  (\bibinfo{year}{2020}).

\bibitem{sadeghiFrequencyFluctuationsNanomechanical2020}
\bibinfo{author}{Sadeghi, P.}, \bibinfo{author}{Demir, A.},
  \bibinfo{author}{Villanueva, L.~G.}, \bibinfo{author}{K{\"a}hler, H.} \&
  \bibinfo{author}{Schmid, S.}
\newblock \bibinfo{title}{Frequency fluctuations in nanomechanical silicon
  nitride string resonators}.
\newblock \emph{\bibinfo{journal}{Physical Review B}}
  \textbf{\bibinfo{volume}{102}}, \bibinfo{pages}{214106}
  (\bibinfo{year}{2020}).

\bibitem{manzanequeResolutionLimitsResonant2023}
\bibinfo{author}{Manzaneque, T.} \emph{et~al.}
\newblock \bibinfo{title}{Resolution {{Limits}} of {{Resonant Sensors}}}.
\newblock \emph{\bibinfo{journal}{Physical Review Applied}}
  \textbf{\bibinfo{volume}{19}}, \bibinfo{pages}{054074}
  (\bibinfo{year}{2023}).

\bibitem{lewisIntroductionFrequencyStandards1991}
\bibinfo{author}{Lewis, L.}
\newblock \bibinfo{title}{An introduction to frequency standards}.
\newblock \emph{\bibinfo{journal}{Proceedings of the IEEE}}
  \textbf{\bibinfo{volume}{79}}, \bibinfo{pages}{927--935}
  (\bibinfo{year}{1991}).

\bibitem{el2020inertial}
\bibinfo{author}{El-Sheimy, N.} \& \bibinfo{author}{Youssef, A.}
\newblock \bibinfo{title}{Inertial sensors technologies for navigation
  applications: State of the art and future trends}.
\newblock \emph{\bibinfo{journal}{Satellite Navigation}}
  \textbf{\bibinfo{volume}{1}}, \bibinfo{pages}{1--21} (\bibinfo{year}{2020}).

\bibitem{amaro2012low}
\bibinfo{author}{Amaro-Seoane, P.} \emph{et~al.}
\newblock \bibinfo{title}{Low-frequency gravitational-wave science with
  elisa/ngo}.
\newblock \emph{\bibinfo{journal}{Classical and Quantum Gravity}}
  \textbf{\bibinfo{volume}{29}}, \bibinfo{pages}{124016}
  (\bibinfo{year}{2012}).

\bibitem{xu2023SiC}
\bibinfo{author}{Xu, M.} \emph{et~al.}
\newblock \bibinfo{title}{High-strength amorphous silicon carbide for
  nanomechanics}.
\newblock \emph{\bibinfo{journal}{Advanced Materials}}
  \textbf{\bibinfo{volume}{36}}, \bibinfo{pages}{2306513}
  (\bibinfo{year}{2024}).

\bibitem{manjeshwar2023InGap}
\bibinfo{author}{Manjeshwar, S.~K.} \emph{et~al.}
\newblock \bibinfo{title}{High-q trampoline resonators from strained
  crystalline ingap for integrated free-space optomechanics}.
\newblock \emph{\bibinfo{journal}{Nano Letters}} \textbf{\bibinfo{volume}{23}},
  \bibinfo{pages}{5076--5082} (\bibinfo{year}{2023}).

\bibitem{pratt_nanoscale_2023}
\bibinfo{author}{Pratt, J.} \emph{et~al.}
\newblock \bibinfo{title}{Nanoscale {Torsional} {Dissipation} {Dilution} for
  {Quantum} {Experiments} and {Precision} {Measurement}}.
\newblock \emph{\bibinfo{journal}{Physical Review X}}
  \textbf{\bibinfo{volume}{13}}, \bibinfo{pages}{011018}
  (\bibinfo{year}{2023}).

\bibitem{gisler2022soft}
\bibinfo{author}{Gisler, T.} \emph{et~al.}
\newblock \bibinfo{title}{Soft-clamped silicon nitride string resonators at
  millikelvin temperatures}.
\newblock \emph{\bibinfo{journal}{Physical Review Letters}}
  \textbf{\bibinfo{volume}{129}}, \bibinfo{pages}{104301}
  (\bibinfo{year}{2022}).

\bibitem{yuan2015silicon}
\bibinfo{author}{Yuan, M.}, \bibinfo{author}{Cohen, M.~A.} \&
  \bibinfo{author}{Steele, G.~A.}
\newblock \bibinfo{title}{Silicon nitride membrane resonators at millikelvin
  temperatures with quality factors exceeding 108}.
\newblock \emph{\bibinfo{journal}{Applied Physics Letters}}
  \textbf{\bibinfo{volume}{107}}, \bibinfo{pages}{263501}
  (\bibinfo{year}{2015}).

\bibitem{balandat2020botorch}
\bibinfo{author}{Balandat, M.} \emph{et~al.}
\newblock \bibinfo{title}{Botorch: A framework for efficient monte-carlo
  bayesian optimization}.
\newblock \emph{\bibinfo{journal}{Advances in neural information processing
  systems}} \textbf{\bibinfo{volume}{33}}, \bibinfo{pages}{21524--21538}
  (\bibinfo{year}{2020}).

\bibitem{mouraCentimeterscaleSuspendedPhotonic2018a}
\bibinfo{author}{Moura, J.~P.}, \bibinfo{author}{Norte, R.~A.},
  \bibinfo{author}{Guo, J.}, \bibinfo{author}{Sch{\"a}fermeier, C.} \&
  \bibinfo{author}{Gr{\"o}blacher, S.}
\newblock \bibinfo{title}{Centimeter-scale suspended photonic crystal mirrors}.
\newblock \emph{\bibinfo{journal}{Optics Express}}
  \textbf{\bibinfo{volume}{26}}, \bibinfo{pages}{1895--1909}
  (\bibinfo{year}{2018}).

\bibitem{maduroMolybdenumNanopillarArrays2021}
\bibinfo{author}{Maduro, L.}, \bibinfo{author}{{de Boer}, C.},
  \bibinfo{author}{Zuiddam, M.}, \bibinfo{author}{Memisevic, E.} \&
  \bibinfo{author}{{Conesa-Boj}, S.}
\newblock \bibinfo{title}{Molybdenum nanopillar arrays: {{Fabrication}} and
  engineering}.
\newblock \emph{\bibinfo{journal}{Physica E: Low-dimensional Systems and
  Nanostructures}} \textbf{\bibinfo{volume}{134}}, \bibinfo{pages}{114903}
  (\bibinfo{year}{2021}).

\bibitem{sainiemiMaskMaterialEffects2007}
\bibinfo{author}{Sainiemi, L.} \& \bibinfo{author}{Franssila, S.}
\newblock \bibinfo{title}{Mask material effects in cryogenic deep reactive ion
  etching}.
\newblock \emph{\bibinfo{journal}{Journal of Vacuum Science \& Technology B:
  Microelectronics and Nanometer Structures Processing, Measurement, and
  Phenomena}} \textbf{\bibinfo{volume}{25}}, \bibinfo{pages}{801--807}
  (\bibinfo{year}{2007}).

\bibitem{reinhardtUltralowNoiseSiNTrampoline2016}
\bibinfo{author}{Reinhardt, C.}, \bibinfo{author}{M{\"u}ller, T.},
  \bibinfo{author}{Bourassa, A.} \& \bibinfo{author}{Sankey, J.~C.}
\newblock \bibinfo{title}{Ultralow-{{Noise SiN Trampoline Resonators}} for
  {{Sensing}} and {{Optomechanics}}}.
\newblock \emph{\bibinfo{journal}{Physical Review X}}
  \textbf{\bibinfo{volume}{6}}, \bibinfo{pages}{021001} (\bibinfo{year}{2016}).

\bibitem{ghadimi2017radiation}
\bibinfo{author}{Ghadimi, A.~H.}, \bibinfo{author}{Wilson, D.~J.} \&
  \bibinfo{author}{Kippenberg, T.~J.}
\newblock \bibinfo{title}{Radiation and internal loss engineering of
  high-stress silicon nitride nanobeams}.
\newblock \emph{\bibinfo{journal}{Nano letters}} \textbf{\bibinfo{volume}{17}},
  \bibinfo{pages}{3501--3505} (\bibinfo{year}{2017}).

\bibitem{serra2018silicon}
\bibinfo{author}{Serra, E.} \emph{et~al.}
\newblock \bibinfo{title}{Silicon nitride moms oscillator for room temperature
  quantum optomechanics}.
\newblock \emph{\bibinfo{journal}{Journal of Microelectromechanical Systems}}
  \textbf{\bibinfo{volume}{27}}, \bibinfo{pages}{1193--1203}
  (\bibinfo{year}{2018}).

\bibitem{reetzAnalysisMembranePhononic2019a}
\bibinfo{author}{Reetz, C.} \emph{et~al.}
\newblock \bibinfo{title}{Analysis of {{Membrane Phononic Crystals}} with
  {{Wide Band Gaps}} and {{Low-Mass Defects}}}.
\newblock \emph{\bibinfo{journal}{Physical Review Applied}}
  \textbf{\bibinfo{volume}{12}}, \bibinfo{pages}{044027}
  (\bibinfo{year}{2019}).

\bibitem{hojUltracoherentNanomechanicalResonators2021b}
\bibinfo{author}{H{\o}j, D.} \emph{et~al.}
\newblock \bibinfo{title}{Ultra-coherent nanomechanical resonators based on
  inverse design}.
\newblock \emph{\bibinfo{journal}{Nature Communications}}
  \textbf{\bibinfo{volume}{12}}, \bibinfo{pages}{5766} (\bibinfo{year}{2021}).

\bibitem{gislerSoftClampedSiliconNitride2022}
\bibinfo{author}{Gisler, T.} \emph{et~al.}
\newblock \bibinfo{title}{Soft-{{Clamped Silicon Nitride String Resonators}} at
  {{Millikelvin Temperatures}}}.
\newblock \emph{\bibinfo{journal}{Physical Review Letters}}
  \textbf{\bibinfo{volume}{129}}, \bibinfo{pages}{104301}
  (\bibinfo{year}{2022}).

\bibitem{guoActivefeedbackQuantumControl2023a}
\bibinfo{author}{Guo, J.}, \bibinfo{author}{Chang, J.}, \bibinfo{author}{Yao,
  X.} \& \bibinfo{author}{Gr{\"o}blacher, S.}
\newblock \bibinfo{title}{Active-feedback quantum control of an integrated
  low-frequency mechanical resonator}.
\newblock \emph{\bibinfo{journal}{Nature Communications}}
  \textbf{\bibinfo{volume}{14}}, \bibinfo{pages}{4721} (\bibinfo{year}{2023}).

\bibitem{schmidFundamentalsNanomechanicalResonators2023}
\bibinfo{author}{Schmid, S.}, \bibinfo{author}{Villanueva, L.~G.} \&
  \bibinfo{author}{Roukes, M.~L.}
\newblock \emph{\bibinfo{title}{Fundamentals of {{Nanomechanical Resonators}}}}
  (\bibinfo{publisher}{{Springer International Publishing}},
  \bibinfo{address}{{Cham}}, \bibinfo{year}{2023}).

\bibitem{verbridgeSizeFrequencyDependent2008}
\bibinfo{author}{Verbridge, S.~S.}, \bibinfo{author}{Ilic, R.},
  \bibinfo{author}{Craighead, H.~G.} \& \bibinfo{author}{Parpia, J.~M.}
\newblock \bibinfo{title}{Size and frequency dependent gas damping of
  nanomechanical resonators}.
\newblock \emph{\bibinfo{journal}{Applied Physics Letters}}
  \textbf{\bibinfo{volume}{93}}, \bibinfo{pages}{013101}
  (\bibinfo{year}{2008}).

\bibitem{doughertyStitchingErrorReductionGratings2001}
\bibinfo{author}{Dougherty, D.~J.}, \bibinfo{author}{Muller, R.~E.},
  \bibinfo{author}{Maker, P.~D.} \& \bibinfo{author}{Forouhar, S.}
\newblock \bibinfo{title}{Stitching-{{Error Reduction}} in {{Gratings}} by
  {{Shot-Shifted Electron-Beam Lithography}}}.
\newblock \emph{\bibinfo{journal}{Journal of Lightwave Technology}}
  \textbf{\bibinfo{volume}{19}}, \bibinfo{pages}{1527} (\bibinfo{year}{2001}).

\bibitem{henryTechniquesCryogenicReactive2009}
\bibinfo{author}{Henry, M.~D.}, \bibinfo{author}{Welch, C.} \&
  \bibinfo{author}{Scherer, A.}
\newblock \bibinfo{title}{Techniques of cryogenic reactive ion etching in
  silicon for fabrication of sensors}.
\newblock \emph{\bibinfo{journal}{Journal of Vacuum Science \& Technology A}}
  \textbf{\bibinfo{volume}{27}}, \bibinfo{pages}{1211--1216}
  (\bibinfo{year}{2009}).

\bibitem{jones1998efficient}
\bibinfo{author}{Jones, D.~R.}, \bibinfo{author}{Schonlau, M.} \&
  \bibinfo{author}{Welch, W.~J.}
\newblock \bibinfo{title}{Efficient global optimization of expensive black-box
  functions}.
\newblock \emph{\bibinfo{journal}{Journal of Global optimization}}
  \textbf{\bibinfo{volume}{13}}, \bibinfo{pages}{455--492}
  (\bibinfo{year}{1998}).

\bibitem{frazier2018tutorial}
\bibinfo{author}{Frazier, P.~I.}
\newblock \bibinfo{title}{A tutorial on bayesian optimization}.
\newblock \emph{\bibinfo{journal}{arXiv preprint arXiv:1807.02811}}
  (\bibinfo{year}{2018}).

\bibitem{NIPS2012_4522}
\bibinfo{author}{Snoek, J.}, \bibinfo{author}{Larochelle, H.} \&
  \bibinfo{author}{Adams, R.~P.}
\newblock \bibinfo{title}{Practical bayesian optimization of machine learning
  algorithms}.
\newblock In \bibinfo{editor}{Pereira, F.}, \bibinfo{editor}{Burges, C. J.~C.},
  \bibinfo{editor}{Bottou, L.} \& \bibinfo{editor}{Weinberger, K.~Q.} (eds.)
  \emph{\bibinfo{booktitle}{Advances in Neural Information Processing Systems
  25}}, \bibinfo{pages}{2951--2959} (\bibinfo{publisher}{Curran Associates,
  Inc.}, \bibinfo{year}{2012}).

\bibitem{shahriari2015taking}
\bibinfo{author}{Shahriari, B.}, \bibinfo{author}{Swersky, K.},
  \bibinfo{author}{Wang, Z.}, \bibinfo{author}{Adams, R.~P.} \&
  \bibinfo{author}{De~Freitas, N.}
\newblock \bibinfo{title}{Taking the human out of the loop: A review of
  bayesian optimization}.
\newblock \emph{\bibinfo{journal}{Proceedings of the IEEE}}
  \textbf{\bibinfo{volume}{104}}, \bibinfo{pages}{148--175}
  (\bibinfo{year}{2015}).

\bibitem{kennedy2000predicting}
\bibinfo{author}{Kennedy, M.~C.} \& \bibinfo{author}{O'Hagan, A.}
\newblock \bibinfo{title}{Predicting the output from a complex computer code
  when fast approximations are available}.
\newblock \emph{\bibinfo{journal}{Biometrika}} \textbf{\bibinfo{volume}{87}},
  \bibinfo{pages}{1--13} (\bibinfo{year}{2000}).

\bibitem{han2012hierarchical}
\bibinfo{author}{Han, Z.-H.} \& \bibinfo{author}{G{\"o}rtz, S.}
\newblock \bibinfo{title}{Hierarchical kriging model for variable-fidelity
  surrogate modeling}.
\newblock \emph{\bibinfo{journal}{AIAA journal}} \textbf{\bibinfo{volume}{50}},
  \bibinfo{pages}{1885--1896} (\bibinfo{year}{2012}).

\bibitem{le2014recursive}
\bibinfo{author}{Le~Gratiet, L.} \& \bibinfo{author}{Garnier, J.}
\newblock \bibinfo{title}{Recursive co-kriging model for design of computer
  experiments with multiple levels of fidelity}.
\newblock \emph{\bibinfo{journal}{International Journal for Uncertainty
  Quantification}} \textbf{\bibinfo{volume}{4}} (\bibinfo{year}{2014}).

\bibitem{williams2007multi}
\bibinfo{author}{Williams, C.}, \bibinfo{author}{Bonilla, E.~V.} \&
  \bibinfo{author}{Chai, K.~M.}
\newblock \bibinfo{title}{Multi-task gaussian process prediction}.
\newblock \emph{\bibinfo{journal}{Advances in neural information processing
  systems}} \bibinfo{pages}{153--160} (\bibinfo{year}{2007}).

\bibitem{liu2018remarks}
\bibinfo{author}{Liu, H.}, \bibinfo{author}{Cai, J.} \& \bibinfo{author}{Ong,
  Y.-S.}
\newblock \bibinfo{title}{Remarks on multi-output gaussian process regression}.
\newblock \emph{\bibinfo{journal}{Knowledge-Based Systems}}
  \textbf{\bibinfo{volume}{144}}, \bibinfo{pages}{102--121}
  (\bibinfo{year}{2018}).

\bibitem{forrester2007multi}
\bibinfo{author}{Forrester, A.~I.}, \bibinfo{author}{S{\'o}bester, A.} \&
  \bibinfo{author}{Keane, A.~J.}
\newblock \bibinfo{title}{Multi-fidelity optimization via surrogate modelling}.
\newblock \emph{\bibinfo{journal}{Proceedings of the royal society a:
  mathematical, physical and engineering sciences}}
  \textbf{\bibinfo{volume}{463}}, \bibinfo{pages}{3251--3269}
  (\bibinfo{year}{2007}).

\bibitem{huang2006sequential}
\bibinfo{author}{Huang, D.}, \bibinfo{author}{Allen, T.~T.},
  \bibinfo{author}{Notz, W.~I.} \& \bibinfo{author}{Miller, R.~A.}
\newblock \bibinfo{title}{Sequential kriging optimization using
  multiple-fidelity evaluations}.
\newblock \emph{\bibinfo{journal}{Structural and Multidisciplinary
  Optimization}} \textbf{\bibinfo{volume}{32}}, \bibinfo{pages}{369--382}
  (\bibinfo{year}{2006}).

\bibitem{jiang2019variable}
\bibinfo{author}{Jiang, P.}, \bibinfo{author}{Cheng, J.},
  \bibinfo{author}{Zhou, Q.}, \bibinfo{author}{Shu, L.} \& \bibinfo{author}{Hu,
  J.}
\newblock \bibinfo{title}{Variable-fidelity lower confidence bounding approach
  for engineering optimization problems with expensive simulations}.
\newblock \emph{\bibinfo{journal}{AIAA Journal}} \textbf{\bibinfo{volume}{57}},
  \bibinfo{pages}{5416--5430} (\bibinfo{year}{2019}).

\bibitem{villanuevaEvidenceSurfaceLoss2014}
\bibinfo{author}{Villanueva, L.~G.} \& \bibinfo{author}{Schmid, S.}
\newblock \bibinfo{title}{Evidence of {{Surface Loss}} as {{Ubiquitous Limiting
  Damping Mechanism}} in {{SiN Micro-}} and {{Nanomechanical Resonators}}}.
\newblock \emph{\bibinfo{journal}{Physical Review Letters}}
  \textbf{\bibinfo{volume}{113}}, \bibinfo{pages}{227201}
  (\bibinfo{year}{2014}).

\end{thebibliography}

\section*{Acknowledgements}

A.C and R.N. acknowledge valuable support from the Kavli Nanolab Delft, in particular from Charles de Boer and Roald van der Kolk. A.C and R. N. would like to thank Matthijs H. J. de Jong and Minxing Xu for stimulating discussions and early assistance with fabrication and experiments. We also thank Tracy Northup for insightful discussions on levitated nanospheres. This work has received funding from the EMPIR programme co-financed by the Participating States and from the European Union’s Horizon 2020 research and innovation programme (No. 17FUN05 PhotoQuant). This publication is part of the project, Probing the physics of exotic superconductors with microchip Casimir experiments (740.018.020) of the research programme NWO Start-up which is partly financed by the Dutch Research Council (NWO). Funded/Co-funded by the European Union (ERC, EARS, 101042855). Views and opinions expressed are however those of the author(s) only and do not necessarily reflect those of the European Union or the European Research Council. Neither the European Union nor the granting authority can be held responsible for them.  D.S., M.A.B and R.N. would like to acknowledge the TU Delft's 3mE Faculty Cohesion grant that enabled to start this project.

\section*{Author contributions}
A.C., D.S., M.A.B., and R.N. designed the research; D.S., L.G., and M.A.B. conducted the data-driven computational design with support from A.C.; A.C. fabricated the nanomechanical resonators with support from R.N., A.C. and R.N led the experiment with support from D.S.; A.C. and D.S. analyzed the data; A.C., D.S., L.G., P.S., M.A.B., and R.N. wrote the paper. A.C. and D.S. contributed equally to this work, as well as M.A.B and R.N.

\section*{Competing interests}
The authors declare no competing interests.

\newpage

\begin{widetext} 

\setcounter{figure}{0}
\renewcommand{\thefigure}{S\arabic{figure}}
\setcounter{equation}{0}
\renewcommand{\theequation}{S\arabic{equation}}
\setcounter{table}{0}
\renewcommand{\thetable}{S\arabic{table}}

\section*{Supplementary information}
\label{SIsec}

\subsection{Gas damping}
\label{SIsec:GasDamping}

 \begin{figure*}[!b] 
	\centering
	\includegraphics[width = 0.85\textwidth]{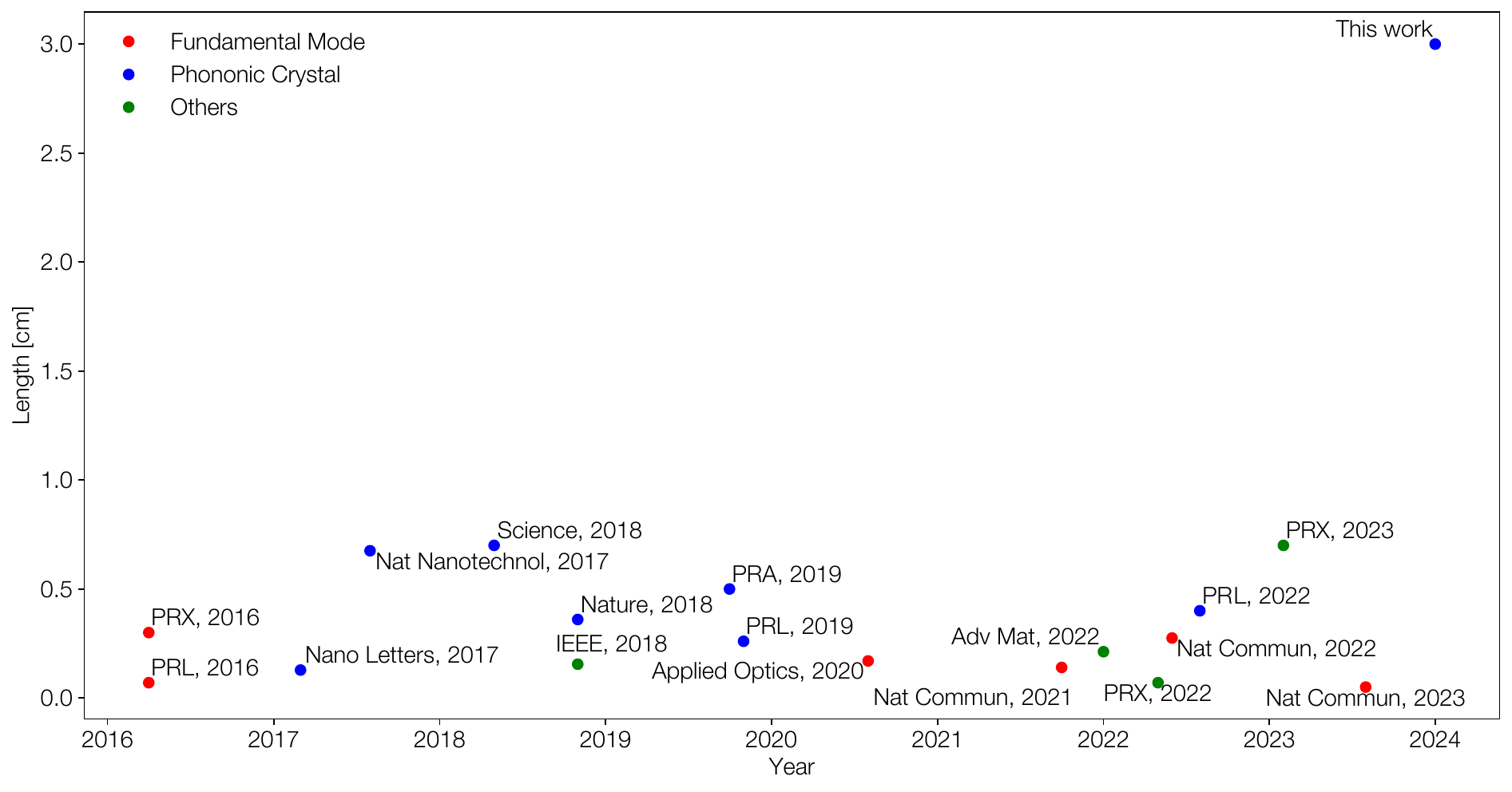} 
	\caption{Room temperature quality factors of resonators fabricated from highly pre-stressed films are proportional to a resonators lateral size. Here we show a variety of high-Q room temperature mechanical resonators over the last years plotted in accordance to their lateral size vs. publication year. The colors indicate what type of mechanical mode is used. The datapoints in the plot, from left to right, correspond to \cite{reinhardtUltralowNoiseSiNTrampoline2016, norteMechanicalResonatorsQuantum2016, ghadimi2017radiation, tsaturyanUltracoherentNanomechanicalResonators2017, ghadimiElasticStrainEngineering2018, rossiMeasurementbasedQuantumControl2018, serra2018silicon, reetzAnalysisMembranePhononic2019a, guoFeedbackCoolingRoom2019, plucharCavityfreeGroundstateCooling2020, hojUltracoherentNanomechanicalResonators2021b, shinSpiderwebNanomechanicalResonators2022, bereyhiPerimeterModesNanomechanical2022, bereyhiHierarchicalTensileStructures2022, gislerSoftClampedSiliconNitride2022, prattNanoscaleTorsionalDissipation2023, guoActivefeedbackQuantumControl2023a}}
	\label{fig:SI0_ReviewerLengths}
\end{figure*}

Room temperature quality factors of resonators fabricated from highly pre-stressed films are typically proportional to the resonator's aspect ratio ($Q \propto L/t$). Thinner films tend to exhibit a lower intrinsic quality, which makes increasing the length (L) a favorable design strategy. If a mechanical mode is only limited by bending losses, designs scaled up in L should show higher quality factor. To utilize this, we strategically aim for the longest possible multi-centimeter scale designs as shown in \autoref{fig:SI0_ReviewerLengths}. Nevertheless, we encounter a trade-off: as resonators grow beyond 3 cm, they are more prone to gas damping — a result of interactions with gas molecules — that increases with lower resonance frequencies, and higher aspect-ratio mechanical modes. Thus longer designs require lower vacuums to evade gas damping \autoref{eq:S5_Q_tot}. With a turbo and roughing pump, the lowest achievable pressure in our UHV chamber is $10^{-9}$~mbar (\autoref{EHARSI_image13_VacuumSetup}), limiting our nanostrings lengths to 3~cm. With more advanced pumping schemes, it is possible to achieve pressures of $10^{-11}$~mbar\cite{daniaUltrahighQualityFactor2023} and further increase resonators' lengths beyond 3~cm to achieve quality factors beyond 10 billion. This section discusses the experiments performed to confirm this boundary.

Gas damping originates by the interaction between the moving surface of the resonators with the gas molecules around it and it often dominates the extrinsic contributions at ambient condition. The nature of this loss mechanism depends on the amount of gas molecules surrounding the resonators, hence the pressure of the gas and the gas composition compared to the resonator's dimension. It can therefore be suppressed by reducing the pressure value at which the resonators operate.

To this end, we first need to evaluate the regime in which the resonator is operating and the consequent dominant gas damping mechanisms. This can be  done by calculating the Knudsen number ($K_n$), describing the ratio of the gas mean free path length ($\lambda_f$) to the representative physical length scale of the resonator ($L_r$) \cite{schmidFundamentalsNanomechanicalResonators2023}: 
\begin{equation}
	\label{eq:S1_Knudsen}
	K_n = \frac{\lambda_f }{L_r}
\end{equation}
The gas mean free path length can then be calculated by the following equation:
\begin{equation}
	\label{eq:S2_lambda_d}
	\lambda_f = \frac{k_B T}{\sqrt{2}\pi d_{gas}^2} \frac{1}{p}
\end{equation}

where $k_B$ is the Boltzmann constant, $T$ is temperature, $ d_{gas}$ is the diameter of the gas particles, and $p$ is the gas pressure. Atmospheric air possesses a mean free path of approximately \SI{70}{nm}, several orders of magnitude lower than the representative physical lengths of the nanomechanical resonators hereby developed. It follows that at the ambient condition the resonators are dominated by viscous damping.

By decreasing the pressure, we enter the ballistic regime where the resonator dimensions become compared or smaller than the gas mean free path (\autoref{eq:S2_lambda_d}), hence $K_n$ becomes larger than unity. In this regime, the quality factor scales with the pressure until gas damping becomes negligible compared to other sources of losses. The quality factor in this regime can  be calculated from an energy transfer model as \cite{verbridgeSizeFrequencyDependent2008}

\begin{equation}
	\label{eq:S3_Qfree}	
	Q_{free} = \frac{\rho t \omega}{p}\sqrt{\frac{\pi RT}{32 M}}
\end{equation}
where $\rho$ is density, $t$ is the resonators thickness, $\omega$ is the resonance frequency, $R$ is the molar gas constant, $T$ is the temperature, and $M$ is the molar mass of the gas (\SI{28.97}{\gram.\mol^{-1}} for air). If there is another surface in close proximity to the resonator, contributions from squeeze-film damping need to be considered and the quality factor can be calculated from

\begin{equation}
	\label{eq:S4_Q_sq}	
	Q_{sq} = 16\pi\frac{d}{L}Q_{free}
\end{equation}
where $d$ is the gap height between the resonator and nearby surface, and $L$ is the resonator length. The total quality factor is then given by 

\begin{equation}
	\label{eq:S5_Q_tot}	
	Q_{tot}(p) = \Big(\frac{1}{Q_{int}}+\frac{c_0}{Q_{sq}(p)}\Big)^{-1}
\end{equation}
where $Q_{int}$ is the intrinsic quality factor when gas damping becomes negligible and $c_0$ is an experimental scaling factor \cite{saarinenLaserCoolingMembraneinthemiddle2023}. It is important to notice that gas damping is not the only source of extrinsic loss. Specifically radiation loss, caused by the hard clamping of the resonator to the supporting substrate, has been reported to play a critical role. However, the nanomechanical resonators here developed employ a PnC to isolate the mechanical mode from environmental noise. This highly suppresses radiation loss as previously reported for similar designs \cite{tsaturyanUltracoherentNanomechanicalResonators2017}. We therefore consider the quality factor limited only by intrinsic contributions once gas damping becomes negligible. The intrinsic losses are addressed in the main text.

To quantify the effect of gas damping we performed ringdown measurements of the 3 cm nanomechanical resonator at different pressure levels and extracted the resulting quality factor from each dataset. The data points are then fitted using \autoref{eq:S5_Q_tot} with $c_0$ and $Q_{int}$ as fitting parameters. The result in Fig. \ref{fig:SI1_GasDamping} demonstrates that gas damping becomes negligible (less than 5\%)  for pressure levels approaching \SI{1e-9}{mbar}. The extracted quality factor equals the intrinsic value of 3.42 billion for this specific resonator. We therefore carried out all the measurements reported in the main text at this pressure level. 

While the specific resonator employed for the pressure study hereby described is not the same device used for the results reported in Fig. 4 of the main text, the geometrical dimensions are equal. We can hence assume a similar gas damping behavior. Moreover \autoref{eq:S5_Q_tot} has been developed for a beam with a uniform width, while the width of our resonators varies along its entire length due to the applied tapering and the PnC. One then has to resort to numerical simulations for an accurate calculation of the quality factor at every pressure level. However, the purpose of this study is to simply find the vacuum requirements to extract the intrinsic quality factor of the fabricated resonators not limited by gas damping, rather than accurately capturing the gas damping behavior. For this purpose, \autoref{eq:S5_Q_tot} provides an accurate lower estimate.
 
 \begin{figure*} 
	\centering
	\includegraphics[width = 0.4\textwidth]{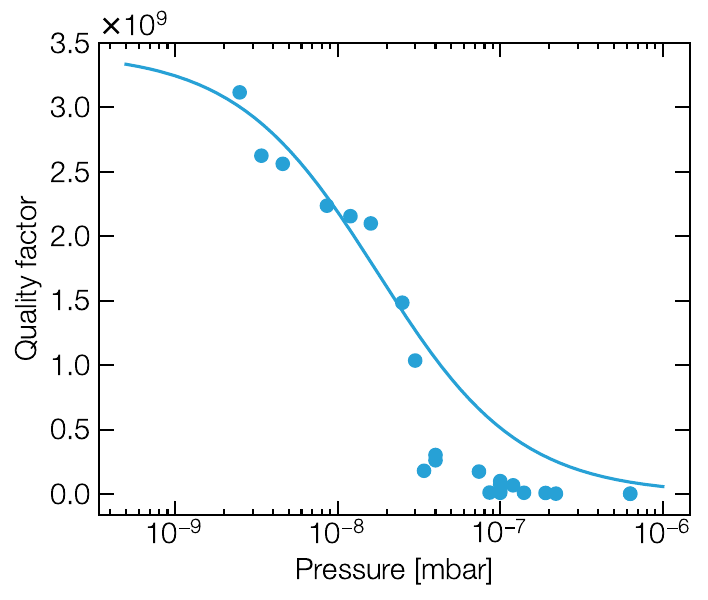} 
	\caption{Measured quality factor as a function of pressure for a 3 cm optimized nanomechanical resonator. The blue line shows the fit, and the blue points indicate the experimental data.}
	\label{fig:SI1_GasDamping}
\end{figure*}

\subsection{Variation of quality factor as function of thickness}
\label{SIsec:Qvsthickness}

The predicted Q factor of the optimized centimeter-scale nanomechanical resonators is 10 billion, while the fabricated structures show a Q factor of 6.6 billion. We believe the observed difference is caused by difficulties to dissipate heat during the undercut process caused by the high-aspect-ratios. This results in different thicknesses and dimensions from edge to center along the beam of the fabricated nanostructures, with a significant effect on the measured Q factor.

In this section, we investigate the case for which the thickness of the Si\textsubscript{3}N\textsubscript{4} layer is uniformly \SI{20}{nm} larger than the expected thickness, a value compatible with the etch settings employed during the undercut in view of the Si\textsubscript{3}N\textsubscript{4} etch rate in SF\textsubscript{6} plasma etching and the temperature of the process.

This thickness variation results in a reduction of the quality factor from over 1 billion at \SI{50}{nm} to \num{6.9e9} at \SI{70}{nm}, as shown in Fig. \ref{fig:SI2_Thickness}a. This value is in good agreement with the value experimentally measured (Fig. 4a of the main text) of \num{6.6e9}. On the contrary, the thickness has a negligible effect on the resonance frequency value as displayed in Fig. \ref{fig:SI2_Thickness}b, where $f_0$ is the resonance frequency obtained for the \SI{30}{nm}-thick resonator, corresponding to \SI{213.323}{kHz}. This value agrees with experiments to around 1\%, further confirming the high-fidelity between simulations and experiments.

 \begin{figure*} 
	\centering
	\includegraphics[width = 0.85\textwidth]{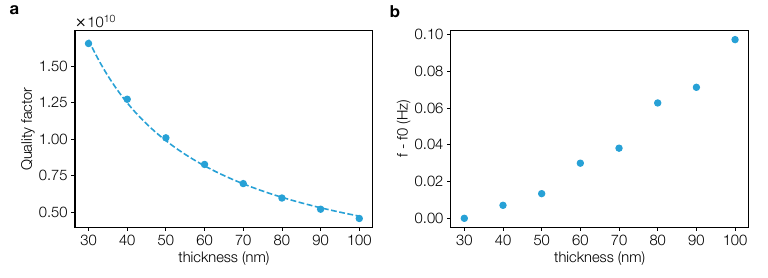} 
	\caption{\textbf{a}, Quality factor as a function of the resonator thickness. \textbf{b}, Difference of the eigenfrequency value as a function of the resonator thickness. The data are obtained by finite element analysis by varying the thickness of the optimized design described in the main text. $f_0$ is the resonance frequency obtained for the \SI{30}{nm}-thick resonator, corresponding to \SI{213.323}{kHz}. }
	\label{fig:SI2_Thickness}
\end{figure*}

\subsection{Stitching errors in electron beam lithography}
\label{SIsec:StitchingErrors}

A pattern in electron beam lithography is generated by deflecting the electron beam based on the desired shape. However, due to the finite sweep range of the deflector system, the writing field cannot exceed an area of typically \SI{100}{\micro m} - \SI{1}{mm}. The maximum writing field for the Raith EBPG 5200 machine at Kavli Nanolab in Delft is \SI{1040}{\micro m}. Large patterns need therefore to be stitched together by moving the sample on a stage. Misalignment between multiple writing fields might then arise due to miscalibration and thermal drift among other reasons. The consequent stitching errors have significant effects on the obtained shape\cite{doughertyStitchingErrorReductionGratings2001}. This is particularly relevant for high stress Si\textsubscript{3}N\textsubscript{4}, where a misalignment of the order of nanometers can lead to stress concentration and rupture of the fragile suspended structure.

The centimeter-scale nanomechanical resonators hereby developed require 30 writing fields to be correctly patterned by electron beam lithography. In order to accurately transfer the desired shape, we employed an electron beam with an estimated spot size equal to \SI{18}{nm} and a spacing of \SI{5}{nm}. The fine resolution results in a long writing time exceeding one hour, which increases the likelihood of misalignment and consequent stitching errors. For most of the exposed devices we in fact observed an incorrect patterning at the boundary of every writing field as shown in the dark-field microscope picture in Fig. \ref{fig:SI3_EBLstiching}a. Figure \ref{fig:SI3_EBLstiching}b provides a picture with higher magnification of the same structure which clearly shows a discontinuity of the written pattern.

In order to mitigate the observed stitching errors we focused first on reducing the long writing time. We did so by performing multiple exposures with different resolutions. The most critical features were exposed with a fine electron beam and high resolution, while a coarse electron beam with lower resolution was employed for the remaining areas of the pattern. As a result, the total writing time for a single \SI{3}{cm} nanomechanical resonator was reduced down to 10 minutes without affecting the accuracy of the desired pattern. This was effective to eliminate the systematic misalignment previously present at every writing field boundary (Fig. \ref{fig:SI3_EBLstiching}a), however, some stitching errors could still be observed in a random manner, varying from run to run. 

To this end, we changed the dose at different positions to take into account possible underdose at the boundary between writing fields and overlapped each writing field with the nearby one for \SI{100}{nm}. This resulted in a correct exposure of the desired pattern, without any noticeable stitching error, and an accurate transfer of the centimeter-scale nanoresonators geometry.

 \begin{figure*} 
	\centering
	\includegraphics[width = 0.85\textwidth]{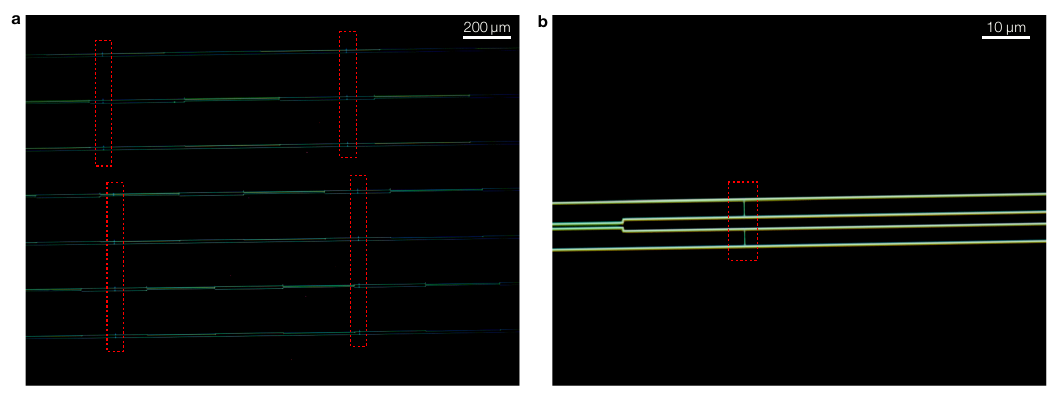} 
	\caption{\textbf{a}, Dark-field microscope picture of a pattern with multiple structures exceeding the writing field of the electron beam lithography. \textbf{b}, Zoom-in dark-field microscope picture of a single structure at the boundary between two writing fields. The dotted red square highlights the misalignment observed between consecutive writing fields. The pictures are taken after developing the exposed e-beam resist.}
	\label{fig:SI3_EBLstiching}
\end{figure*}

\subsection{Photoresist cracking at cryogenic temperature}
\label{SIsec:PhotoresistCracking}

Cryogenic deep reactive ion etching (DRIE) is employed in the fabrication of the centimeter-scale nanoresonators to increase the gap separating the suspended  Si\textsubscript{3}N\textsubscript{4} from the supporting Si substrate. Contrary to other deep silicon etching techniques as the Bosch process, cryo-DRIE is known to result in a smoother sidewall \cite{henryTechniquesCryogenicReactive2009}, and it does not leave carbon residues on the sidewall. In fact, the oxide passivation layer created to anisotropically etch Si dissolves at high temperatures. At the same time, its main drawback is the vulnerability of photoresists to cracking at low temperatures.

Previous studies found that cracking depends on the layer thickness and the specific photoresist material. The thin photoresist of less than \SI{1.5}{\micro m} are in fact reported to be free from cracking. Moreover, materials with a high degree of cross-linking and high mechanical strength do not suffer from cracking even for larger thickness. We then studied the effect of cryo-DRIE on the photoresist SU1813, spin coated on Si\textsubscript{3}N\textsubscript{4} chips with thickness varying from \SI{1.4}{\micro m} to \SI{2}{\micro m}, before being exposed to the fluorine-based cryo-DRIE. The thickness is varied within an interval which provides enough material to obtain the desired gap size due to the photoresist selectivity. The thicker photoresist (\SI{2}{\micro m}) shows cracks over the entire surface (Fig. \ref{fig:SI4_PhotoresistCracks}c), while in the thinner photoresist (\SI{1.4}{\micro m}) the cracks are localized at the outer area of the chip (Fig. \ref{fig:SI4_PhotoresistCracks}b), in good agreement with the expected behavior. The cracks for the outer area of the thin photoresists are most likely caused by the non-uniform thickness at the edge, as Fig. \ref{fig:SI4_PhotoresistCracks}a shows. We repeated the analysis for a second photoresist, AZ5214, without noticing any significant difference.

 \begin{figure*} 
	\centering
	\includegraphics[width = 0.85\textwidth]{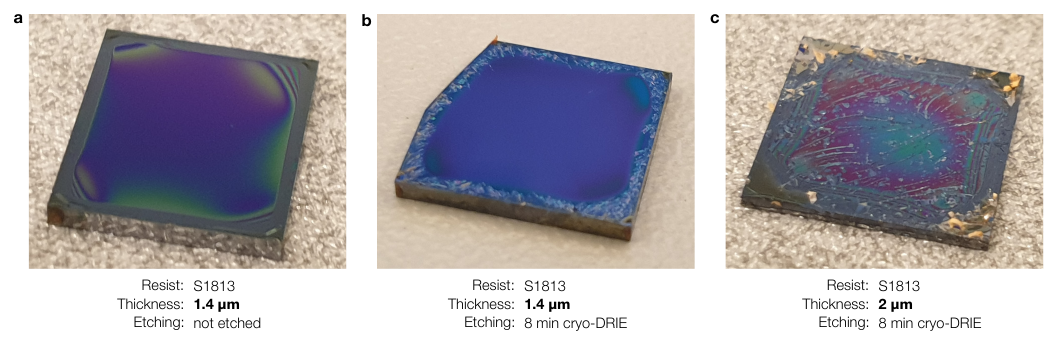} 
	\caption{Photograph of silicon chip with photoresist spin coated on top before and after cryo-DRIE etching. \textbf{a}, Device before the cryo-DRIE step with uniform photoresist layer \textbf{b, c}, Devices with different photoresist layer after cryo-DRIE etching.}
	\label{fig:SI4_PhotoresistCracks}
\end{figure*}

We therefore focused on a photoresist thickness of \SI{1.4}{\micro m} (Fig. \ref{fig:SI4_PhotoresistCracks}b) in which the central area remains intact. Despite most of the surface being free from cracking, we observed that some cracks might propagate from the outer area towards the center with fatal consequences on the nanoresonators. Figure \ref{fig:SI5_PhotoresistCracks2}a shows an example where the cracks propagate toward the nanomechanical resonators area. The cracked resist patterns can then transfer into the underneath Si\textsubscript{3}N\textsubscript{4} during etching, destroying the devices. A post-baking step of 15 minutes at \SI{120}{\celsius} prior to the etching was found to be effective to improve the photoresist plasma resistance and to further to reduce the cracks on the surface, but not to completely remove them. We also investigated the effect of the carrier wafer thickness and the Si substrate thickness. Previous studies found in fact that photoresist cracking is initiated by wafer deformation caused by the helium backside cooling system and can be mitigated by employing a thicker substrate \cite{sainiemiMaskMaterialEffects2007}. However, in our case, the photoresist cracking did not show any dependence on the substrate thickness.

 \begin{figure*} 
	\centering
	\includegraphics[width = 0.85\textwidth]{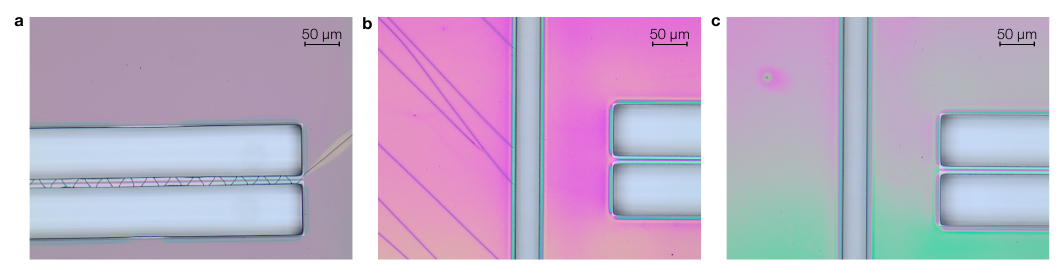} 
	\caption{Optical microscope pictures of the photoresist layer after cryo-DRIE. \textbf{a}, Pattern without outer ring in which the cracks propagate toward the centimeter-scale nanoresonator. \textbf{b-c}, Pattern with the outer ring which protects the centimeter-scale nanoresonator from the cracks.}
	\label{fig:SI5_PhotoresistCracks2}
\end{figure*}

We therefore surrounded the nanomechanical resonator's area of the device with a protecting outer ring (Fig. \ref{fig:SI6_PhotoresistCracks3}). This ring is etched into both the  Si\textsubscript{3}N\textsubscript{4} layer and the upper part of Si substrate in order to create a physical barrier. The resulting pattern is shown in \ref{fig:SI5_PhotoresistCracks2}b which clearly shows the effectiveness of the developed method to prevent any cracks from propagating toward the nanomechanical resonators. It is nevertheless important to notice that often photoresist cracking was confined to the outer area of the device without propagating toward the central area (\ref{fig:SI5_PhotoresistCracks2}c),

 \begin{figure*} 
	\centering
	\includegraphics[width = 0.4\textwidth]{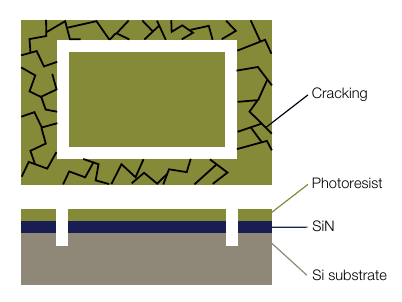} 
	\caption{Schematic describing the protecting outer design.}
	\label{fig:SI6_PhotoresistCracks3}
\end{figure*}

\subsection{Beams collapsing}
\label{SIsec:BeamCollpase}

The centimeter-scale nanomechanical resonators are suspended by a fluorine based plasma etching performed at cryogenic temperature. The etching allows a quick and controllable release of the Si\textsubscript{3}N\textsubscript{4} layer, without limitations from surface tension. However, plasma etching introduces charging effects due to the insulating nature of Si\textsubscript{3}N\textsubscript{4}. The charged Si\textsubscript{3}N\textsubscript{4} can then be attracted by the Si substrate leading to the collapse of the nanoresonators.

Depending on the distance between the suspended resonator and the surrounding, we observed the Si\textsubscript{3}N\textsubscript{4} being attracted toward or the side (Fig. \ref{fig:SI7_Collpase}c) or the bottom (Fig. \ref{fig:SI7_Collpase}b) of the opening around it. It follows that controlling the gap distance is an effective way to mitigate the collapse of the structures.

The distance from the wall along the in-plane direction can easily be controlled lithographically during the electron beam exposure. We observed that an opening larger than \SI{70}{\micro m} is needed to eliminate the probability of sticking to the wall. This affects the exposed area and thus the total writing time, which in turn can increase the probability of stitching errors. However as discussed in section \ref{SIsec:StitchingErrors}, a multi exposure with different resolutions is an effective solution to reduce the total writing time and eliminate stitching errors.

On the other hand, the distance between the suspended resonator and the Si substrate along the out-of-plane direction needs to be controlled by an etching step. To this end, we performed a cryo-DRIE of the Si substrate prior to the release step. A gap under the suspended Si\textsubscript{3}N\textsubscript{4} larger than \SI{50}{\micro m} was observed to be needed to avoid the collapse of the structure. The final pattern used in the main text has then a distance between the resonator and the wall of around \SI{80}{\micro m}, while the gap under the Si\textsubscript{3}N\textsubscript{4} after the cryo-DRIE is of \SI{70}{\micro m} (Fig. \ref{fig:SI7_Collpase}a) .

 \begin{figure*} 
	\centering
	\includegraphics[width =0.85\textwidth]{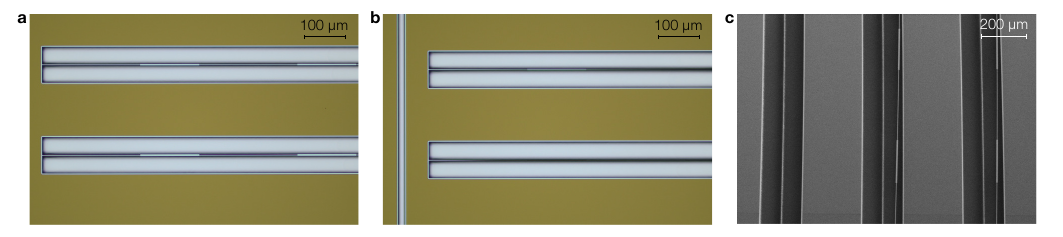} 
	\caption{\textbf{a}, Optical microscope picture of 3 cm nanomechanical resonators correctly suspended. \textbf{b}, Optical microscope picture of 3 cm nanomechanical resonators collapsed. \textbf{c}, Scanning electron microscope picture of 3 cm nanomechanical resonators sticking to the walls.}
	\label{fig:SI7_Collpase}
\end{figure*}

\subsection{Optothermal effects on the measured Q factor}
\label{SIsec:PhototermalEffect}

The quality factor of the fabricated centimeter-scale resonators is experimentally measured with the ringdown method. The resonator is mechanically excited close to resonance with a sinusoidal function before turning off the excitation and measuring the decay time. A linear fit in the logarithmic scale of the observed decay allows us to extract the decay rate and thus the quality factor. It is however of paramount importance to avoid any unintentional excitation of the resonator during the measurement to extract the correct value of the quality factor.

One possible source of unintentional excitation is the laser employed to interferometrically probe the displacement of the resonator. The optical power impinging on the suspended resonator can in fact optothermally or optically drive it. To address this issue, we employed a laser operating at \SI{1550}{nm}, the wavelength at which Si\textsubscript{3}N\textsubscript{4} has negligible absorption. Moreover, we minimized the laser power coupling to the resonator down to \SI{400}{nW}. This value is several orders of magnitude lower than the laser power conventionally used in previous studies (see for example \cite{shinSpiderwebNanomechanicalResonators2022}), where no effects correlated to optothermal or optical excitation were observed. To further dismiss any contributions we performed different ringdown measurements by varying the laser power incident on the resonator from \SI{40}{\micro W} to \SI{400}{nW}. The results in Fig. \ref{fig:SI8_RingdownLaser} show a comparable decay rate for all the traces, suggesting that optothermal or optical effects are indeed negligible. The measurements are performed on the optimized design reported in the main text in Fig. 4.

 \begin{figure*} 
	\centering
	\includegraphics[width =0.4\textwidth]{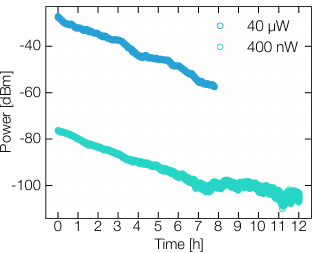} 
	\caption{Ringdown traces of the optimized nanoresonator excited at \SI{214}{kHz}. The motion is probed with different laser powers in consecutive measurements, plotted in the same graph for comparison.}
	\label{fig:SI8_RingdownLaser}
\end{figure*}

\subsection{Multi-fidelity Bayesian optimization}
\label{SIsec:MFBO}

The pursuit of efficient optimization of data scarce and high-fidelity black-box functions has led to Bayesian optimization techniques. These are statistical methods that yield a belief model over the entire domain $\mathcal{X}$ which is sequentially updated through newly acquired data. The generic Bayesian optimization (BO) algorithm was originally introduced by Jones \textit{et al.} \cite{jones1998efficient}, and is a proxy-optimization scheme: instead of optimizing $f$ directly over $\mathcal{X}$, one first selects a regressor $R$ to model the response surface $\hat{f}(\mathcal{X})$ based on a design of experiments $\mathcal{D}=\{(x_1,f(x_1)),\ldots,(x_n,f(x_n))\}\subset\mathcal{X}$, which is simply a set of $n$ known input-output pairs. Importantly, the response surface also includes a measure of uncertainty on top of the predicted outcome. This is why BO is often performed with Gaussian process regression (GPR) \cite{frazier2018tutorial, NIPS2012_4522, shahriari2015taking}. Based on this regression model $R(\mathcal{D})$, a so-called acquisition function is built and optimized over the same domain. The goal of this proxy optimization is to suggest a new point $x_{n+1}$ in $\mathcal{X}$ to be sampled with $f$, and as such, the design of experiments is augmented with $(x_{n+1},f(x_{n+1}))$. In an algorithmic format, this can be expressed as follows:
\begin{algorithm}[H]
\caption{Bayesian optimization}
\begin{algorithmic}[1]
\Require{Search space $\mathcal{X}$, regressor $R$, design of experiments $\mathcal{D}$, acquisition function acq, threshold condition $C$}
\While{$C$ is False}\\
$\hat{f}\leftarrow R(\mathcal{D})$\\
$x\leftarrow\mathrm{argmax}_{x'\in\mathcal{X}}\text{acq}(x';\hat{f})$\\
$y\leftarrow f(x)$\\
$\mathcal{D}\leftarrow\mathcal{D}\cup\{(x,y)\}$
\EndWhile
\end{algorithmic}
\label{sfbo-algo}
\end{algorithm}
Given the scarcity of high-fidelity data, a standard solution is to acquire higher-throughput data with lower fidelity, i.e., with larger uncertainty with lower cost. This context of simultaneous high- and low-fidelity data structures, combined with the ideas of GPR, gives rise to a multi-fidelity data driven modelling paradigm. The first multi-fidelity GPR (MFGPR) method was introduced by Kennedy \& O'Hagan \cite{kennedy2000predicting}, under the term cokriging. Numerous other MFGPR methods have been constructed and researched since then \cite{han2012hierarchical, le2014recursive, williams2007multi}, reviewed by Liu \textit{et al}. \cite{liu2018remarks}.

There are several ways in which Bayesian optimization (Algorithm \ref{sfbo-algo}) can be extended to handle regressors $R$ over data sets with multiple fidelities, as is the case with MFGPR. A straightforward way can be described as follows, when the \textit{m} indicates the function's fidelity:
\begin{algorithm} [H]
\caption{Multi-fidelity Bayesian optimization}
\begin{algorithmic}[1]
\Require{Search space $\mathcal{X}$, fidelity space $\mathcal{Z}$, regressor $R$, DoE $\mathcal{D}$, acquisition function acq, threshold condition $C$}
\While{$C$ is False}\\
$\hat{f}\leftarrow R(\mathcal{D})$\\
$x,m\leftarrow\mathrm{argmax}_{(x',m')\in\mathcal{X}\times\mathcal{Z}}\text{acq}(x',m';\hat{f})$\\
$y\leftarrow f_m(x,m)$\\
$\mathcal{D}\leftarrow\mathcal{D}\cup\{(x,y)\}$
\EndWhile
\end{algorithmic}
\label{smfbo}
\end{algorithm}
Note that this algorithm updates Algorithm \ref{sfbo-algo}, so that the acquisition function samples from both the fidelity function.

Interfacing MFGPR with BO has been discussed in practice by Forrester \textit{et al}. \cite{forrester2007multi} and Huang \textit{et al}. \cite{huang2006sequential} While the former simply applied expected improvement (EI) acquisition on the prediction of a single-fidelity, the latter devised an augmented version of EI. This acquisition function multiplies the EI function applied to the highest fidelity, with fidelity-dependent parameters such as the ratio of computational cost and correlation between the highest fidelity and the fidelity in question. This effectively separates the design space and fidelity space aspects of the multi-fidelity problem. More recently, Jiang et al. \cite{jiang2019variable} have applied a similar multiplicative factor principle to create variable-fidelity upper confidence bound (VFUCB).
\\\\
To demonstrate the process outlined by Algorithm \ref{smfbo}, we use cokriging as the regressor $(R)$ along with VFUCB as the 2-fidelity ($\mathcal{Z}=\{\text{low},\text{high}\}$) acquisition function (acq). See Fig. \ref{fig:acq} for visualization in the case of multi-fidelity data $\mathcal{D}$ sampled from the high- and low-fidelity Forrester functions \cite{forrester2007multi}, a set of similar one-dimensional objective functions.

\begin{figure}[H]
    \centering
    \includegraphics[width=.35\textwidth]{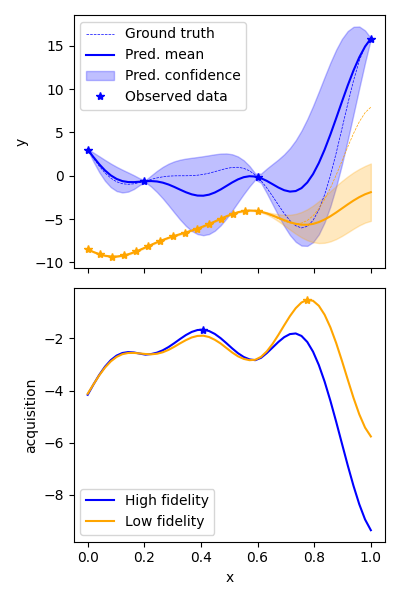}
    \caption{(top) The input-output space of the objective function. The blue solid line indicates the cokriging predictive mean for the high-fidelity data (blue stars) sampled from the high-fidelity objective function (blue dashed line). The blue shaded area corresponds to the confidence interval (two predictive standard deviations). The orange counterparts show similar results for the low-fidelity scenario. (bottom) The input-acquisition value space. The blue and orange solid lines indicate the high- and low-fidelity branches of the VFUCB acquisition function respectively. The blue and orange stars respectively show the locations at which the high- and low-fidelity acquisition branches are maximized.}
    \label{fig:acq}
\end{figure}
From Figure \ref{fig:acq}, the following can be inferred:
\begin{itemize}
    \item The maximum acquisition value of the low-fidelity branch is higher than that of the high-fidelity acquisition. Therefore, the fidelity selection $m=\text{high}$ is made in step 3 of Algorithm \ref{smfbo}.
    \item Compared to the maximizer of the high-fidelity acquisition branch, the maximizing $x$-value of the low-fidelity branch is closer to the minimizing $x$-value of the high-fidelity objective; the low-fidelity data is able to guide the optimization process.
\end{itemize}

\subsection{Multi-fidelity Bayesian optimization initial random points dependency}
\label{SIsec:Random}

\begin{figure}
  \centering
    \includegraphics[width=0.8\textwidth]{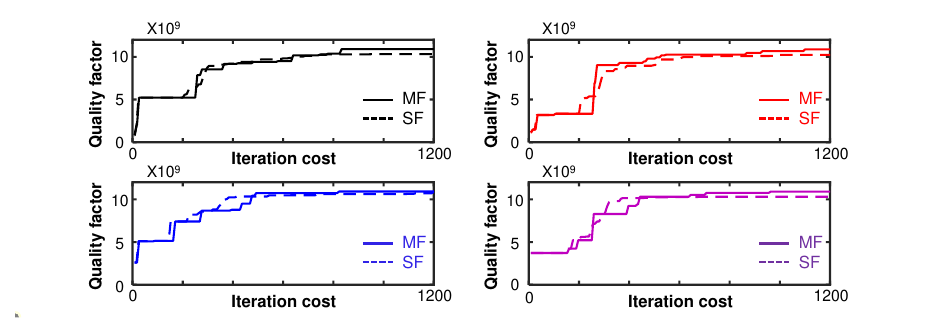}
  \caption{Iteration history of the PnC beam nanomechanical resonator for optimizing $Q_m$ with four different randomly selected initial points. The top-left is the iteration history discussed in the main text.}
  \label{EHARSI_image9_Initialrandom}
\end{figure}

Since our simulation-based design problem handles a stochastic optimization on the beam-like nanomechanical resonator, the initial design of experiments affects the convergence to the optimum solution. The resonator has a design space with nine design parameters, making the initial points ($=$25) affect the performance. Because we use multi-fidelity Bayesian optimization to reduce the total number of simulation evaluations, the curse of dimension is inevitable. \textbf{Figure \ref{EHARSI_image9_Initialrandom}} shows the optimization history when we start from four different sets of random initial points. The change of initial points affects the convergence speed as expected, along with different behavior in selecting the fidelity. However, the optimized design obtained for all cases has converged on a similar design by taking advantage of the multi-fidelity optimization.

\subsection{PnC beam resonator's design parameters}
\label{SIsec:DesignParam}

\begin{figure}
  \centering
    \includegraphics[width=0.8\textwidth]{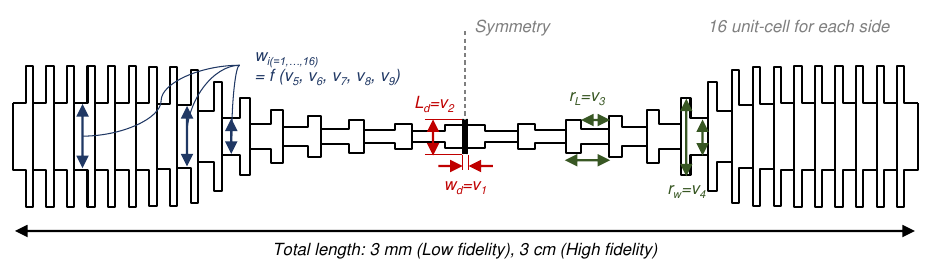}
  \caption{Nanomechanical resonator model with 16 unit cells for each side and 9 design parameters. The total length is 3 cm (high-fidelity) and 3 mm (low-fidelity).}
  \label{EHARSI_image10_Designparam}
\end{figure}

The one-dimensional PnC resonator has a total length of \SI{3}{cm} (or \SI{3}{mm}) length with a total of 32 unit cells.
During the Bayesian optimization, the quality factor of the resonator is maximized by changing nine design parameters.
Figure \ref{EHARSI_image10_Designparam} illustrates the design parameters of the optimized resonator discussed in the manuscript.
Two parameters correspond to the defect's width ($w_d$) and length ($L_d$) in the bound of [\SI{0.5}{\micro m}, \SI{3}{\micro m}], [\SI{150}{\micro m}, \SI{1000}{\micro m}] ([\SI{15}{\micro m}, \SI{100}{\micro m}]), respectively.
Two parameters illustrate the width ($r_w$) and the length($r_L$) ratio for each of the unit-cell in the bound of $[1.5, 3]$, $[0.25, 0.75]$, respectively.
The width ratio is the ratio between the wide and thin parts of the unit cells, and the length ratio is the ratio between the length of the thin part of the unit cell versus the total length of the unit cells.
The tapered shape was defined by five design parameters in the bound of [\SI{0.5}{\micro m}, \SI{3}{\micro m}].
We performed Piecewise Cubic Hermite Interpolating Polynomial for the 16 unit-cell's width of the thin part on the shape-determinating design parameters.
The simulation was performed to find the maximum quality factor in the range of 100 kHz to 400 kHz, considering the defect mode using the bandgap. 
The length of the unit cells was determined considering the bandgap frequency matching condition, once the set of each unit cell's width is defined\cite{ghadimiElasticStrainEngineering2018}.
During the optimization, the resonator's thickness was set to 50 nm.

\subsection{Quality factors of additional devices}
This section presents the measurements conducted on additional fabricated devices, which share the same optimized design and are lithographically identical to the device featured in Fig. 4 of the main text. Figure \ref{EHARSI_image12_RingdownDevices} shows multiple ringdown traces from 3 devices. The uppermost blue curves correspond to four different measurements of the device discussed in the main text, while the green and light purple curves depict measurements obtained from two additional devices, each measured twice. Traces for the same device exhibit a comparable decay rate, resulting in the same quality factor, which is indicated above the first curve of each device. Linear fits for each trace are depicted as solid black lines. 

The power values on the y-axis are adjusted relative to the maximum value of each curve before applying a 10 dBm offset between them. An additional 10 dBm spacing along the y-axis distinguishes different devices. Strong fluctuations resulting from unwanted temperature drifts and mechanical vibrations of the setup can push the measured signal outside the linear region of the interference signal for a brief time interval. This, in turn, leads to occasional spikes, as visible in some curves.
The results demonstrate a consistent and reproducible quality factor among different measurements of each device. The observed variation among the quality factor of different devices suggests differences in surface quality among the devices, e.g., surface impurities and surface roughness \cite{villanuevaEvidenceSurfaceLoss2014}, or different thicknesses. As further detailed in Sec. B, variation in the final thickness of each device might occur due to the difficulties in dissipating heat during the undercut process, significantly impacting the measured quality factor. Among all the devices fabricated and measured, we observed a variation in quality factor of around one order of magnitude, which is in line with typical variations found in other works.

\begin{figure}
    \includegraphics[]{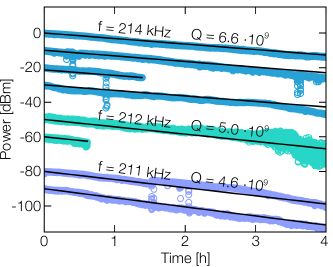}
    \centering
    \caption{Multiple ringdown traces of 3 fabricated nanoresonators with the optimized design presented in the main text. Traces corresponding to measurements from the same device are color-coded for easy identification. To facilitate comparison, each trace has been vertically shifted relative to its respective maximum value, ($P(t)-P_{max}$). Subsequently, an incremental 10 dBm offset has been applied to each trace, starting with the top-most blue trace. An additional 10 dBm offset is applied between measurements from different devices.}
    \label{EHARSI_image12_RingdownDevices}
\end{figure}

\subsection{Ultra High Vacuum Setup}

We developed a UHV setup designed to operate at pressures close to \SI{e-9}{mbar}, aiming to minimize gas damping and measure the intrinsic quality factor of the fabricated centimeter-scale resonators. This pressure value approaches the final pressure of the employed turbomolecular pump, necessitating setup optimization to reduce the volume and the surface of the overall vacuum system, and potential outgassing sources. This section provides details on the vacuum setup.

\begin{figure}[!b]
    \includegraphics[width=0.7\textwidth]{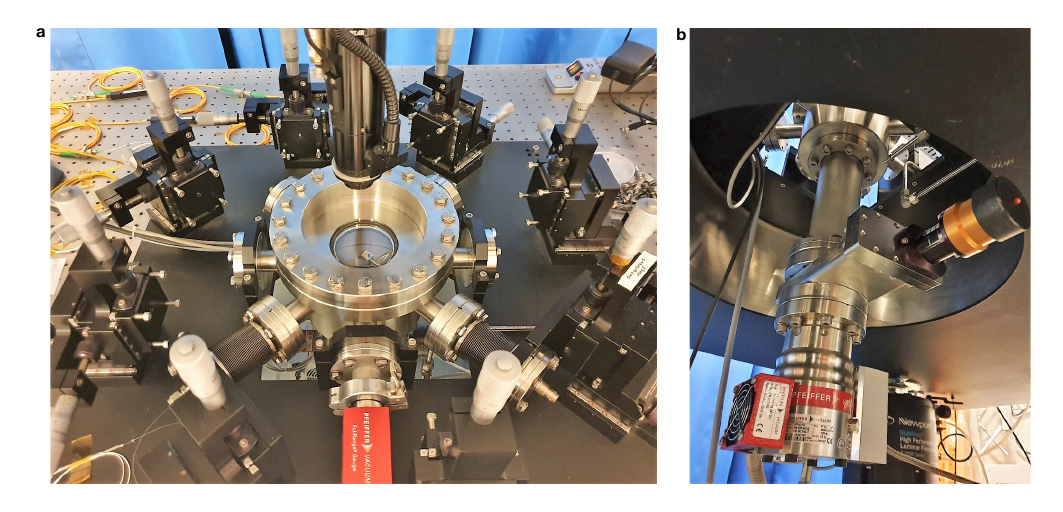}
    \centering
    \caption{Vacuum chamber and turbo molecular pump. \textbf{a}, Photograph of the vacuum chamber with the pressure gauge mounted at the vacuum chamber. \textbf{b}, Photograph of the turbo molecular pump connected to the bottom of the vacuum chamber.}
    \label{EHARSI_image13_VacuumSetup}
\end{figure}

The vacuum setup consists of a cylindrical chamber (Fig. \ref{EHARSI_image13_VacuumSetup}a) with 8 CF-40 flanges along the lateral surface and 2 CF-100 flanges on the top and the bottom. The CF-40 flanges are employed for electrical and optical connection and the vacuum gauge. The bottom flange directly connects to a turbo molecular pump via a valve and a large opening CF-63 (Fig. \ref{EHARSI_image13_VacuumSetup} b) without any restrictive hoses that could hinder pumping efficiency. This allows the volume and the surface of the overall vacuum system to be reduced, improving the pumping efficiency. The chamber incorporates UHV-compatible components for sample manipulation, including a triaxial nanopositioner, sample holders, and piezoelectric plates soldered with lead-free solder to KAPTON wires. Lacking a load lock, we fully open the chamber for each new sample through the top viewport and do not require chamber baking. After loading a sample, we typically purge the chamber with nitrogen three times while the backing pump is running before starting the turbo molecular pump.

It is crucial to note that the employed pressure gauge's operational range spans from $10^{-9}$ to $10^3$ mbar, whereas the turbomolecular pump's final pressure is \SI{5e-10}{mbar}. Consequently, the pressure gauge provides only a lower estimate of the vacuum level.

\end{widetext}

\end{document}